\documentclass[5p,times]{elsarticle}

\usepackage{amsmath,amssymb,amsfonts}
\usepackage{algorithmic}
\usepackage{graphicx}
\usepackage{textcomp}
\usepackage{xcolor}
\usepackage{booktabs}
\usepackage{multirow}
\usepackage{hyperref}
\usepackage{listings}
\usepackage[T1]{fontenc}
\usepackage{inconsolata}
\usepackage[caption=false]{subfig}
\usepackage{verbatim}

\definecolor{belizehole}{HTML}{2980b9}
\definecolor{clouds}{HTML}{f4f8f9}
\definecolor{midnightblue}{HTML}{2c3e50}
\definecolor{pomegranate}{HTML}{c0392b}
\definecolor{royalblue}{HTML}{3867d6}
\definecolor{greensea}{HTML}{16a085}
\definecolor{nephritis}{HTML}{27ae60}
\definecolor{amethyst}{HTML}{9b59b6}

\lstdefinestyle{interfaces}{
	float=t,
	floatplacement=t,
}

\journal{Parallel Computing}

\begin{document}

\begin{frontmatter}



\title{Accelerating Communication for Parallel Programming Models on GPU Systems}


\author[aff1]{Jaemin Choi}
\author[aff1]{Zane Fink}
\author[aff1]{Sam White}
\author[aff2]{Nitin Bhat}
\author[aff3]{David F.~Richards}
\author[aff1,aff2]{Laxmikant V.~Kale}

\affiliation[aff1]{organization={Department of Computer Science, University of Illinois at Urbana-Champaign},
            state={Illinois},
            country={USA}}
\affiliation[aff2]{organization={Charmworks, Inc.},
            state={Illinois},
            country={USA}}
\affiliation[aff3]{organization={Center for Applied Scientific Computing, Lawrence Livermore National Laboratory},
            state={California},
            country={USA}}

\begin{abstract}
As an increasing number of leadership-class systems embrace GPU accelerators in the race towards exascale, efficient communication of GPU data is becoming one of the most critical components of high-performance computing. For developers of parallel programming models, implementing support for GPU-aware communication using native APIs for GPUs such as CUDA can be a daunting task as it requires considerable effort with little guarantee of performance. In this work, we demonstrate the capability of the Unified Communication X (UCX) framework to compose a GPU-aware communication layer that serves multiple parallel programming models of the Charm++ ecosystem: Charm++, Adaptive MPI (AMPI), and Charm4py. We demonstrate the performance impact of our designs with microbenchmarks adapted from the OSU benchmark suite, obtaining improvements in latency of up to 10.1x in Charm++, 11.7x in AMPI, and 17.4x in Charm4py. We also observe increases in bandwidth of up to 10.1x in Charm++, 10x in AMPI, and 10.5x in Charm4py. We show the potential impact of our designs on real-world applications by evaluating a proxy application for the Jacobi iterative method, improving the communication performance by up to 12.4x in Charm++, 12.8x in AMPI, and 19.7x in Charm4py.
\end{abstract}




\end{frontmatter}


\section{Introduction}

The parallel processing power of GPUs has become central to the performance
of today's High Performance Computing (HPC) systems, with seven of the top ten supercomputers
in the world equipped with GPUs~\cite{top500}.
GPU-accelerated applications often store the bulk of their data in device memory,
increasing the importance of efficient inter-GPU data transfers on modern systems.

Although vendors provide GPU programming models such as CUDA for executing
kernels and transferring data,
their limited functionality makes it challenging to implement a general
communication backend for parallel programming models on distributed-memory machines.
Direct GPU-GPU communication crossing the process boundary can be implemented using
CUDA Inter-process Communication (IPC), but requires extensive optimizations
such as IPC handle cache and pre-allocated device buffers~\cite{ipdpsw12-cuda_ipc}.
Direct inter-node transfers
of GPU data cannot be implemented solely with CUDA and requires additional support
from the networking stack~\cite{icpp13-potluri}.
Adding support for GPUs from other vendors such as AMD or Intel requires
more development and optimization efforts that could be spent elsewhere.

A number of software frameworks, such as GASNet~\cite{gasnet}, libfabric~\cite{hoti15-libfabric}, and UCX~\cite{ucx_paper}, aim to provide a unified
communication layer over diverse networking hardware.
While a few have been successfully adopted in parallel programming models
including MPI and PGAS,
UCX is arguably the first communication framework to support
production-grade, high-performance inter-GPU communication on a wide range of
modern GPUs and interconnects.
In this work, we take advantage of UCX's capability to perform direct GPU-GPU transfers
to support GPU-aware communication in multiple parallel programming models
from the Charm++ ecosystem: Charm++, Adaptive MPI (AMPI), and Charm4py.
We extend the UCX machine layer in the Charm++ runtime system to enable
the transfer of GPU buffers and expose this functionality to the parallel programming models,
with model-specific implementations to support their user applications.
Our tests on a leadership-class system show that this approach substantially improves the performance
of GPU-aware communication for all models.

The major contributions of this work are the following:
\begin{itemize}
	\item We present designs and implementation details to enable GPU-aware communication using
	UCX as a common abstraction layer in multiple parallel programming models: Charm++, AMPI, and Charm4py.
	\item We discuss design considerations in the GPU Messaging API to support message-driven execution and task-based runtime systems.
	\item We present the Channel API in Charm++ to improve GPU-aware communication performance by avoiding overheads from metadata messages required in the GPU Messaging API.
	\item We demonstrate the performance impact of our mechanisms using a set of microbenchmarks
	and a proxy application representative of a scientific workload.
\end{itemize}


\section{Background}\label{sec:background}

\subsection{GPU-aware Communication}

GPU-aware communication has developed out of the need to rectify
productivity and performance issues with data transfers involving
GPU buffers. Without GPU-awareness, additional code is required
to explicitly move data between host and device memory,
which also substantially increases latency and reduces attainable bandwidth.

The GPUDirect~\cite{gpudirect} family of technologies have been
leading the effort to resolve such issues on NVIDIA GPUs.
Version~1.0 allows Network Interface Controllers (NICs) to have shared
access to pinned system memory with the GPU and avoid
unnecessary memory copies, and version~2.0 (GPUDirect P2P) enables direct memory access and data transfers between GPU devices on the same PCIe bus.
GPUDirect RDMA~\cite{gpudirect_rdma} utilizes Remote Direct Memory Access (RDMA) technology to allow the NIC to directly access memory on the GPU.
Based on GPUDirect RDMA, the GDRCopy library~\cite{gdrcopy_paper} provides an efficient low-latency transport for small messages.
The Inter-Process Communication (IPC) feature introduced in CUDA~4.1 enables direct transfers between GPU data mapped to different processes,
improving the performance of communication crossing the process boundary~\cite{ipdpsw12-cuda_ipc}.

MPI is one of the first parallel programming models and communication standards to adopt these technologies
and support GPUs in the form of CUDA-aware MPI,
which is available in most MPI implementations. Other parallel programming models have added support for GPU-aware communication
by either implementing their own mechanisms with GPUDirect and CUDA IPC or adopting a communication library such as UCX.

\subsection{UCX}

Unified Communication X (UCX)~\cite{ucx_paper} is an open-source,
high-performance communication framework that provides abstractions over
various networking hardware and drivers, including TCP, OpenFabrics Alliance (OFA)
verbs, Intel Omni-Path, and Cray uGNI. It is currently being developed at a fast pace
with contributions from multiple hardware vendors as well as the open-source
community.

UCX provides support for tag-matching send/receive, stream-oriented send/receive, active messages, remote memory access, and atomic operations.
UCP is the high-level protocol layer in UCX, whose API can be used by parallel programming models to implement a performance-portable communication backend.
Projects using UCX include Dask, OpenMPI, MPICH, and Charm++.
GPU-aware communication is supported on NVIDIA and AMD GPUs through its tagged and stream APIs.
When provided with pointers to GPU memory, these APIs utilize the respective CUDA or ROCm libraries to perform efficient GPU-GPU transfers.

\begin{figure}
\centering
\includegraphics[width=\linewidth]{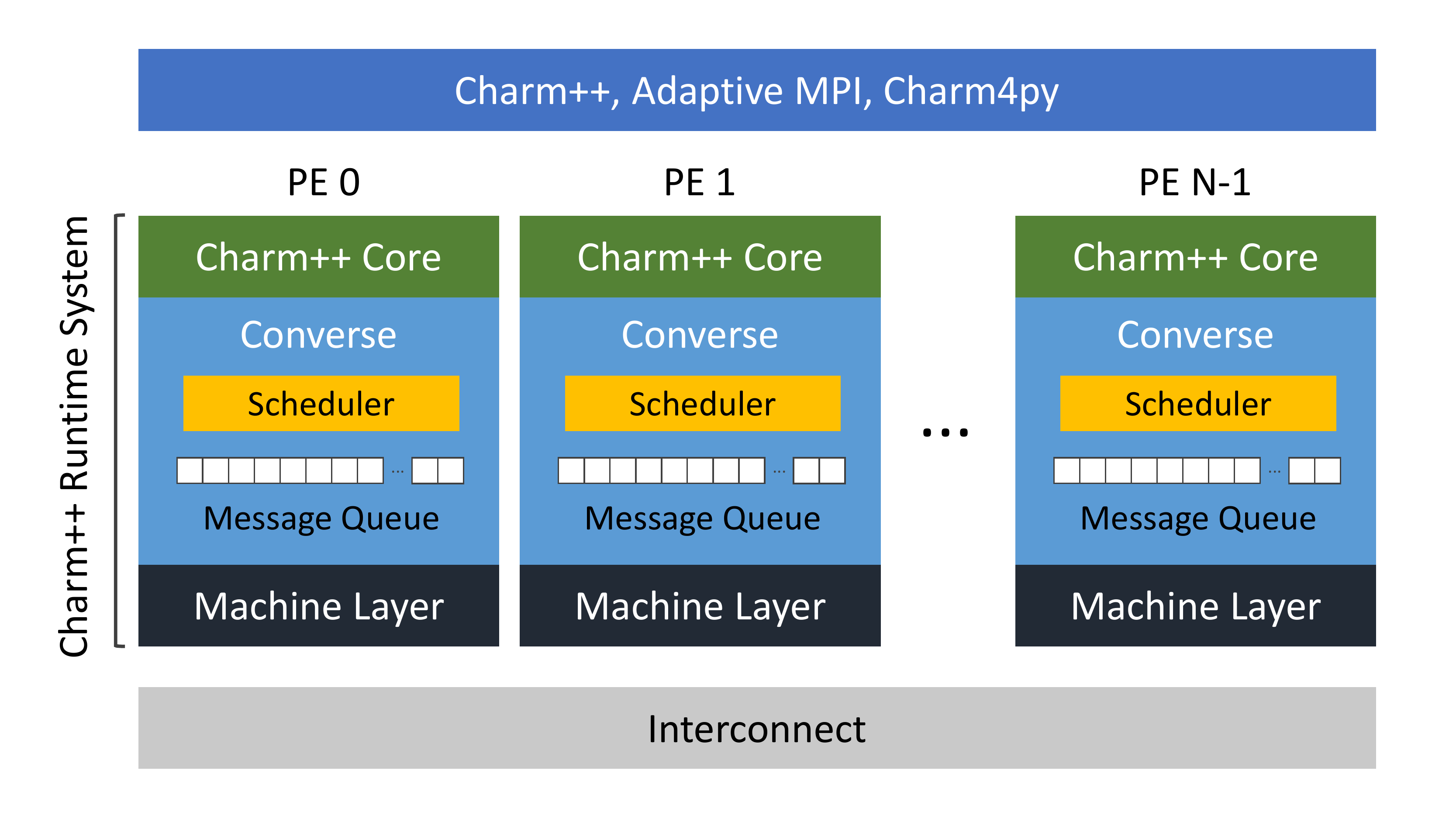}
\caption{Software stack of the Charm++ family of parallel programming models.}
\label{fig:charm_layers}
\vspace{-10pt}
\end{figure}

\subsection{Charm++}

Charm++~\cite{charm} is a parallel programming system based on the C++ language, developed around the concept of migratable objects.
A Charm++ program is decomposed into objects called \textit{chares} that execute in parallel on the Processing Elements (PEs, typically CPU cores), which are scheduled by the underlying runtime system.
This object-centric approach enables \textit{overdecomposition}, where the problem domain can be decomposed into a larger
number of chares than the number of available PEs.
Overdecomposition empowers the runtime system to control the mapping and
scheduling of chares onto PEs, facilitating computation-communication overlap and dynamic load balancing.

The execution of a Charm++ program is driven by messages exchanged between chare objects. The arrival of a message initiates
some work to be performed on the receiver chare object, which can be seen as an execution model based on active messages~\cite{isca92-active_messages}.
Each message encapsulates information about the work to be performed on the receiver chare, i.e., \textit{entry method},
and the data needed for its execution.
Figure~\ref{fig:charm_layers} depicts the structure of the Charm++ Runtime System (RTS) and how it interacts with the
above parallel programming models and the underlying network hardware.
Incoming messages are stored in a message queue maintained per PE, which are eventually picked up by the scheduler.
The message queue and scheduler are part of a software layer called Converse.
The communication mechanism in Charm++ is entirely asynchronous as the sender does not wait for any reply or acknowledgment
from the receiver, and incoming messages are stored in a message queue without blocking the receiver.
Communication operations initiated by chare objects are ultimately handled by the \textit{machine layer}, which interfaces with
the underlying interconnect.
Charm++ supports various low-level transports through its machine layers, including TCP/IP, Mellanox Infiniband, Cray uGNI, IBM PAMI, and UCX.

Charm++ has had limited support for GPU-GPU transfers implemented
with CUDA memory copies and IPC, which only works on a single node with inadequate performance.
With this work, Charm++ now supports GPU-aware communication seamlessly within and across nodes
using UCX, improving the performance of GPU-accelerated applications
developed with any of the parallel programming models in the Charm+ ecosystem.

\subsection{Adaptive MPI}
Adaptive MPI (AMPI)~\cite{lcpc03-ampi} is an MPI library implementation developed on top of the Charm++ runtime system.
AMPI virtualizes the concept of an MPI rank: whereas a traditional MPI library equates ranks with operating system processes, AMPI supports execution with multiple ranks per process by associating each rank with a chare object.
This empowers AMPI to co-schedule ranks that are located on the same PE based on the delivery of messages.
Users can tune the number of ranks they run with based on performance.
AMPI ranks are also migratable at runtime for the purposes of dynamic load balancing or checkpoint/restart-based fault tolerance.

Communication in AMPI is handled through Charm++ and its optimized networking layers.
AMPI optimizes communication based on locality of the recipient rank as well as the size and datatype of the message buffer.
Small buffers are packed inside a regular Charm++ message in an eager fashion,
and the Zero Copy API~\cite{charm_zerocopy} is used to implement a rendezvous protocol for larger buffers.
The underlying runtime optimizes message transmission based on locality over user-space shared memory,
Cross Memory Attach (CMA) for within-node, or RDMA across nodes. This work extends such optimizations
to the context of multi-GPU nodes connected by a high performance network programmable with UCX.

\subsection{Charm4Py}\label{sec:background_charm4py}

Charm4Py~\cite{cluster18-charm4py} is a parallel programming framework based on the Python language,
developed on top of the Charm++ runtime system. It seeks to provide an easily-accessible parallel programming
environment with improved programmer productivity through Python, while maintaining high scalability and performance
of the adaptive C++-based runtime. Being based on Python, Charm4py can readily take
advantage of many widely-used software libraries such as NumPy, SciPy, and pandas.

Chare objects in Charm4py communicate with each other by asynchronously invoking entry methods like Charm++.
The parameters are serialized and packed into a message that is handled by the underlying Charm++ runtime system.
This allows our extension of the UCX machine layer to also support Charm4py.
Charm4py also provides a functionality to establish streamed connections between a pair of chares, called \textit{Channels}~\cite{charm4py_channels}.
Channels provide explicit send/receive semantics to exchange messages similar to MPI,
but retains asynchrony by suspending the caller object until communication is complete.
We extend the Channels feature to support GPU-aware communication in Charm4py, which is discussed in Section~\ref{sec:design_charm4py}.

\section{Design and Implementation}\label{sec:design}

To accelerate communication of GPU data, we utilize the
capability of UCX to directly send and receive GPU data.
UCX is supported as a machine layer in Charm++,
positioned at the lowest level of the software stack directly interfacing
the interconnect, as illustrated in Figure~\ref{fig:charm_layers}.
As AMPI and Charm4py are also built on top of the Charm++ RTS,
all host-side communication travels through the various layers of the RTS
where layer-specific headers are added or extracted,
with actual communication primitives executed by the machine layer.

\begin{figure}[t]
\centering
\begin{lstlisting}[language=C++]
// Sender object's method
void Sender::foo() {
	// Send a message to the receiver object
	// to execute the 'bar' entry method
	receiver.bar(my_val1, my_val2);
}

// Receiver object's entry method,
// executed once the sender's message
// is picked up by the scheduler
void Receiver::bar(int val1, double val2) {
	// val1 and val2 are available
	...
}
\end{lstlisting}
\caption{Message-driven execution in Charm++.}
\label{fig:charm_entry_method}
\vspace{-10pt}
\end{figure}

The message-driven execution model in Charm++, as shown in Figure~\ref{fig:charm_entry_method},
necessitates additional metadata to be attached to each message, so that the receiver
can figure out how to handle the message. This involves determining which chare object
and entry method a message is targeting. Charm++ in traditional CPU-based systems
achieves this by allocating a message that is big enough for both the metadata and
user's payload on host memory, copying the payload into the message, and passing
the prepared message to the machine layer to be sent. On GPU systems, however,
this becomes a challenging problem since the user's payload data can now be in GPU memory
while the metadata still needs to be in host memory (because the RTS is running on the CPU).

Our first approach of supporting GPU-aware communication in the Charm++ RTS,
the \textit{GPU Messaging API}, maintains the message-driven execution model.
This is achieved by retaining the host-side message that contains
the necessary metadata and user's data in host memory, while separately sending the user's data in GPU memory
through the UCX machine layer. Once the host-side message arrives on the receiver (which is possible
with pre-posted receives for the message queue), a receive for the incoming GPU data can be posted using information extracted
from the metadata in the host-side message. The destination GPU buffer is provided by the user
using a mechanism that builds on the \textit{Zero Copy API}~\cite{charm_zerocopy}, which is explained
in more detail in Section~\ref{sec:charm_gpu_messaging}.
A noticeable limitation of the GPU Messaging API is the delay in
posting the receive for the GPU data, as it can only be performed
after the arrival of the host-side message.

To further improve performance from the GPU Messaging API, we have developed the \textit{Channel API} in Charm++.
It allows a `channel' to be created between a pair of chare objects, which exposes two-sided send and receive
semantics that directly translate into communication primitives in UCX.
This eliminates the overhead from the host-side message needed by the GPU Messaging API, but it can
be seen as a deviation from the message-driven execution model; only data moves between chares, not
the flow of execution. This may make it difficult for some applications (especially irregular applications
that benefit from a message-driven model) to adopt the Channel API.
The implementation of the Channel API is described in Section~\ref{sec:charm_channel}.
While Charm++ supports both the GPU Messaging API and the Channel API, GPU-aware communication in
AMPI and Charm4py currently only utilize the GPU Messaging API. Integration of the Channel API
into AMPI and Charm4py is being worked on as it requires extensive changes in the respective codebases
to move away from the message-driven communication model.

In the following sections, we discuss the updates to the UCX machine layer and the various parallel programming models
in the Charm++ ecosystem to enable GPU-aware communication.

\subsection{UCX Machine Layer}

Originally contributed by Mellanox, the UCX machine layer in Charm++ is designed
to handle low-level communication on networks supported by UCX. It utilizes the Tagged API
exposed by the UCP layer in UCX, enabling messages to be exchanged with two-sided communication routines.
Although the UCP Active Message API is potentially a better fit for Charm++, it was presumably not
production-ready at the time of the UCX machine layer implementation and it also currently lacks support for GPUs.

During the initialization phase of the Charm++ runtime, each process
creates a UCP worker and establishes endpoints between all workers
using PMIx~\cite{pc18-pmix}. In the non-SMP mode of Charm++ where each PE
is contained in a single process, there is a UCP worker for each PE.
In the SMP mode, PEs are implemented using threads and multiple PEs can be
contained in each process; the UCP worker in each process is managed by
a separate communication thread. In this work, the non-SMP mode of
Charm++ is used.

Once the endpoints are established,
UCP tagged receives for eager messages (\texttt{ucp\_tag\_recv\_nb}) are posted in advance
to handle incoming messages. All eager host-side communication uses the same tag, \texttt{UCX\_MSG\_TAG\_EAGER},
allowing the UCX machine layer to receive messages from any other PE.
This conforms to the message-driven execution model in Charm++, since the runtime system
must handle incoming messages without relying on receives issued by the user application.
In addition to the pre-posted receives for eager messages, the UCP worker also probes for
rendezvous messages (\texttt{ucp\_tag\_probe\_nb}) with another tag, \texttt{UCX\_MSG\_TAG\_PROBE}.
When a probe is successful, the message is received with \texttt{ucp\_tag\_msg\_recv\_nb}.
The invocation of an entry method from a chare object creates a message, which travels down to the machine layer.
It is then sent to the target endpoint using UCP tagged sends (\texttt{ucp\_tag\_send\_nb}),
with \texttt{UCX\_MSG\_TAG\_EAGER} for eager and \texttt{UCX\_MSG\_TAG\_PROBE} for rendezvous messages.
The non-blocking send and receive UCP calls are advanced by the progress function, \texttt{ucp\_worker\_progress},
which is executed by the Charm++ scheduler. When the UCX machine layer receives a message,
it is stored in the corresponding PE's message queue to be picked up by the scheduler and execute the target entry method.

\begin{figure}
\centering
\subfloat[][GPU Messaging API]{\includegraphics[width=.6\linewidth]{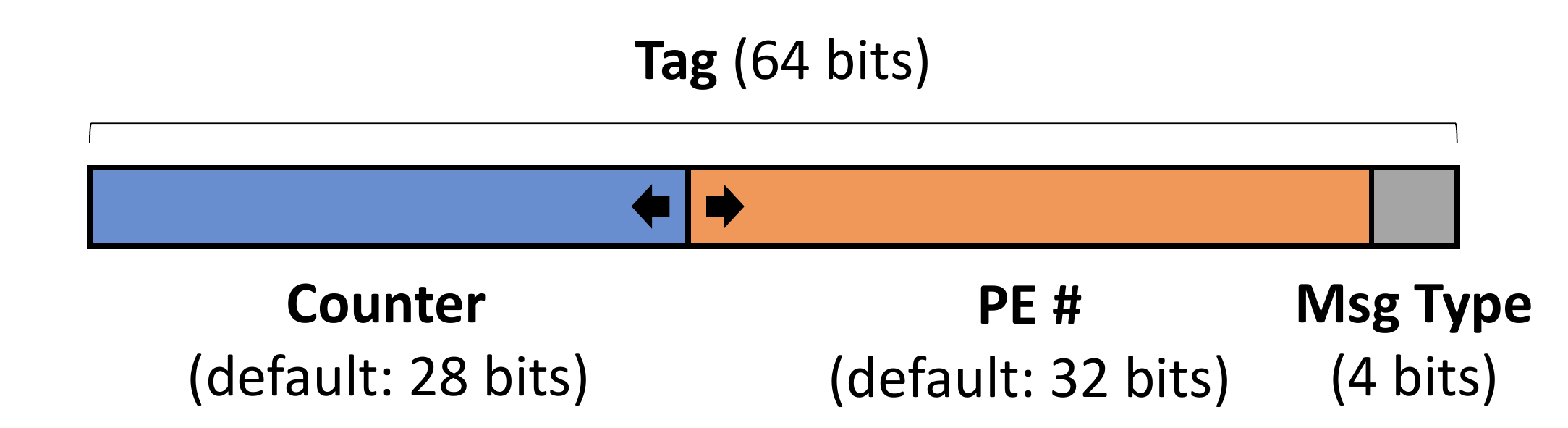}\label{fig:ucx_tag_gpu_messaging}}
\\
\subfloat[][Channel API]{\includegraphics[width=.6\linewidth]{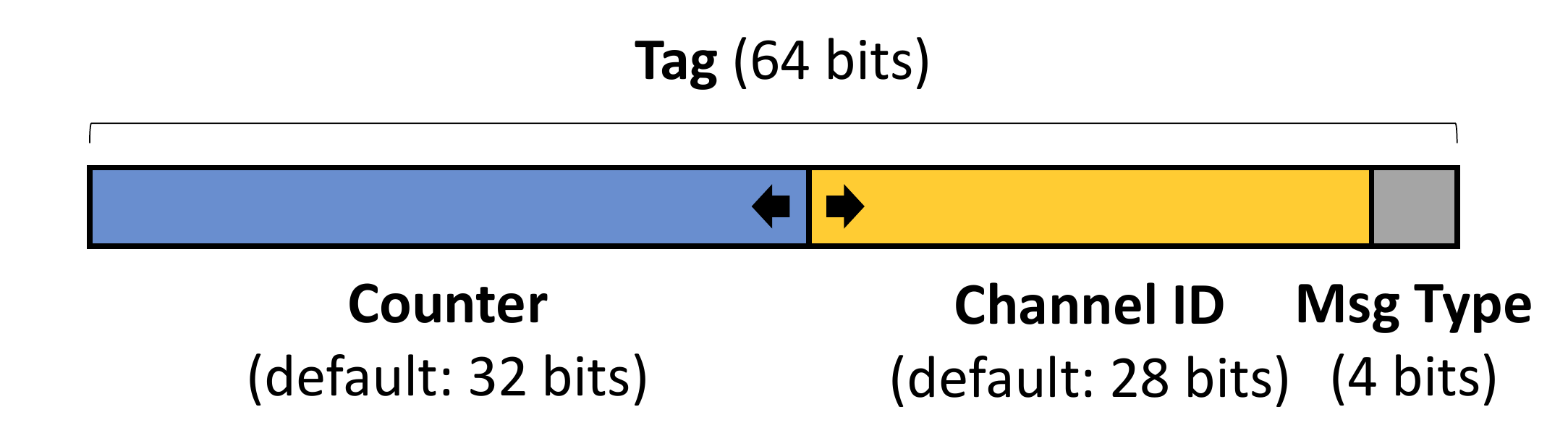}\label{fig:ucx_tag_channel}}
\caption{Tag generation schemes to support GPU-aware communication in the UCX machine layer.}
\label{fig:ucx_tag}
\vspace{-10pt}
\end{figure}

To support GPU-aware communication in the Charm++ family of parallel programming models,
we extend the UCX machine layer to provide an additional interface for
sending and receiving data in GPU memory with the UCP Tagged API.
We have added support for both the GPU Messaging API and Channel API,
with the former used by all three programming models (Charm++, AMPI, and Charm4py)
whereas the latter is currently only used in Charm++. Since both APIs build on
the two-sided send and receive semantics, we implement different schemes to generate
tags for passing to the underlying UCX communication primitives as shown in Figure~\ref{fig:ucx_tag}.

For the GPU Messaging API, the first four bits of the 64-bit tag are used to store
the message type, which is set to a new value called \texttt{UCX\_MSG\_TAG\_DEVICE}.
This allows the Charm++ RTS to recognize the arrival of a host-side metadata message
used by the GPU Messaging API, as described in Section~\ref{sec:charm_gpu_messaging}.
The remainder of the tag is split into storing the source PE index (32 bits by default)
and the value of the counter maintained by the source PE (28 bits by default)
which is incremented on each transmission of GPU data. This division can be modified at compile
time to accommodate different scenarios.

With the Channel API, tag counters are not maintained by the source PE but by
each endpoint chare of the channel. Each chare increments its counter when a send or
receive call is made to the channel.
A new message type, \texttt{UCX\_MSG\_TAG\_CHANNEL}, is assigned to the first four bits
of the 64-bit tag to distinguish messages used by the Channel API.
By default, 28 bits are used to store the channel ID, which
must be unique in the program. The remaining 32 bits are retrieved from the per-channel
counters maintained by the sender or receiver chare participating in the channel.

The core functionalities of GPU-aware communication in the UCX machine layer
are exposed to the upper layers with the functions below, and their usage
is described in the following sections.
\begin{lstlisting}[language=C++, morekeywords={size_t,uint64_t}]
// GPU Messaging API
void LrtsDeviceSend(int dest_pe, const void*& ptr,
                    size_t size, uint64_t& tag);
void LrtsDeviceRecv(DeviceRdmaOp* op, DeviceRecvType type);

// Channel API
void LrtsChannelSend(int dest_pe,
                     const void*& ptr, size_t size,
                     void* cb, uint64_t tag);
void LrtsChannelRecv(const void*& ptr, size_t size,
                     void* cb, uint64_t tag);
\end{lstlisting}

\subsection{Charm++}

Two different mechanisms have been implemented to support GPU-aware communication in the
Charm++ runtime system: (1) GPU Messaging API and (2) Channel API. We discuss why and
how these mechanisms are designed, along with their implications on performance.

\subsubsection{GPU Messaging API}\label{sec:charm_gpu_messaging}

\begin{figure}[t]
\centering
\begin{lstlisting}[language=C++, morekeywords={chare,entry,device,size_t}]
// Charm++ Interface (CI) file
// Declares chare objects and their entry methods
chare MyChare {
	entry MyChare();
	entry void recv(device char data[size], size_t size);
};
\end{lstlisting}
\begin{lstlisting}[language=C++, morekeywords={chare,entry,device,size_t}]
// C++ source file
// (1) Sender chare
void MyChare::send() {
	peer.recv(CkDeviceBuffer(send_gpu_data), size);
}

// (2) Receiver's post entry method
void MyChare::recv(char*& data, size_t& size) {
	// Set the destination GPU buffer
	// Receive size is optional
	data = recv_gpu_data;
}

// (3) Receiver's regular entry method
void MyChare::recv(char* data, size_t size) {
	// Receive complete, GPU data is available
	...
}
\end{lstlisting}
\begin{lstlisting}[language=C++, morekeywords={size_t, uint64_t}]
// Converse layer metadata
struct CmiDeviceBuffer {
	const void* ptr; // Source GPU buffer address
	size_t size;
	uint64_t tag; // Set in the UCX machine layer
	...
};

// Charm++ core layer metadata
struct CkDeviceBuffer : CmiDeviceBuffer {
	CkCallback cb; // Support Charm++ callbacks
	...
};
\end{lstlisting}
\caption{Example Usage of the GPU Messaging API in Charm++.}
\label{fig:charm_gpu_messaging_code}
\vspace{-10pt}
\end{figure}

\begin{figure}[t]
\centering
\includegraphics[width=\linewidth]{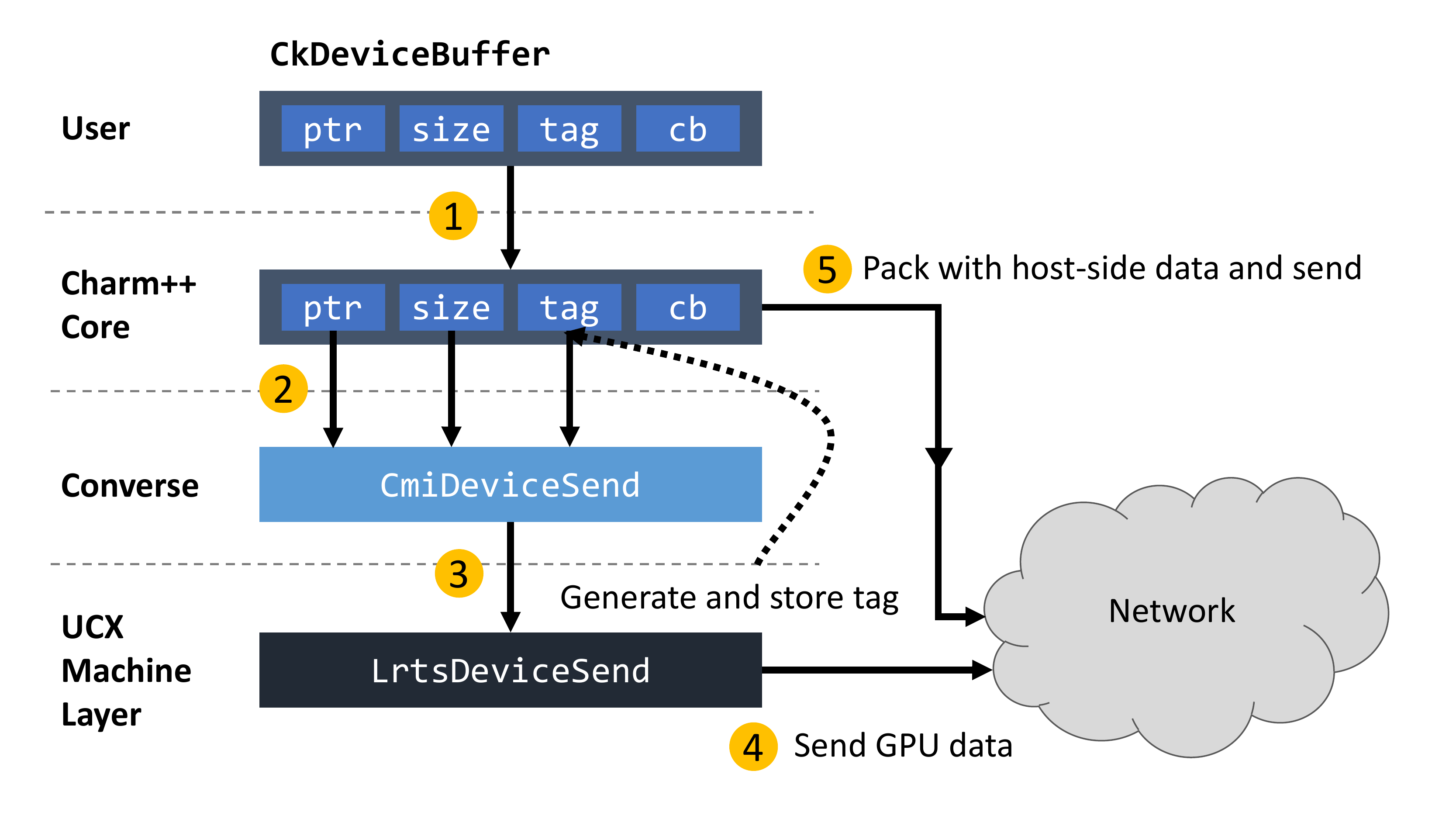}
\caption{Sender-side logic of the GPU Messaging API in Charm++.}
\label{fig:charm_gpu_messaging_design}
\vspace{-10pt}
\end{figure}

Taking inspiration from the Zero Copy API, the GPU Messaging API retains the entry method syntax and message-driven execution model,
where the flow of execution is transferred from the sender chare to the receiver.
As shown in Figure~\ref{fig:charm_gpu_messaging_code}, we provide an additional attribute,
\texttt{device}, to allow users to annotate GPU buffers in the Charm++ Interface (CI) file.
The CI file contains declarations of chare objects and their entry methods, and potentially
the overall structure of parallel execution which are used for code generation.
An entry method invocation such as \texttt{peer.recv} executes a generated code block,
which is modified to send both the source GPU buffer and a separate host-side message that contains
information about the GPU data transfer. This information is encapsulated in a structure
called \texttt{CkDeviceBuffer}, which is passed to the Charm++ core layer and then down to
the Converse layer in the form of \texttt{CmiDeviceBuffer}, then finally to the UCX machine layer
to perform the send of GPU data with \texttt{LrtsDeviceSend}. Charm++ callbacks, stored as \texttt{cb}
in \texttt{CKDeviceBuffer}, are used to notify the sender or receiver that the GPU data transfer is complete.
This process is illustrated in Figure~\ref{fig:charm_gpu_messaging_design}.

The 64-bit tag stored in \texttt{CmiDeviceBuffer} (and by inheritance in \texttt{CkDeviceBuffer})
is set to the correct value by the UCX machine layer in the \texttt{LrtsDeviceSend} function.
It is used in sending the GPU data, and is also transferred to the receiver
chare inside the host-side message, to be used as the tag for receiving the incoming GPU data.
Once the host-side message arrives on the destination PE, the corresponding receive
for the incoming GPU data is posted in \texttt{LrtsDeviceRecv}.
The \texttt{DeviceRdmaOp} struct stores and maintains information necessary for the receive
operations in the various layers of the Charm++ RTS, including the address of the destination
GPU buffer, size of the data, and the tag set by the sender. \texttt{DeviceRecvType}
denotes which of the parallel programming models (Charm++, AMPI, or Charm4py) has posted
the receive, allowing the appropriate handler function to be invoked once the GPU data has been received.

To receive the incoming GPU data directly into the user's destination buffer
and avoid an extra copy, we provide a \textit{post entry method} that allows the user
to specify the address of the destination GPU buffer in advance.
An example usage is shown in Figure~\ref{fig:charm_gpu_messaging_code}.
The post entry method is executed by the runtime system when the host-side message associated
with the GPU data transfer arrives. Once the RTS is informed of the destination GPU buffer address,
it posts a receive for the incoming GPU data with \texttt{LrtsDeviceRecv}, also using information
contained in the host-side message such as the tag used in the UCP send.
Once the GPU data arrives, the regular entry method of the receiver chare
is executed, at which point the received GPU buffer is available to the user.
One downside of the GPU Messaging API is that performance may degrade from the delay
in posting the receive for the incoming GPU data, which arises from the receiver
not knowing which UCX tag was used until the host-side message arrives.

\subsubsection{Channel API}\label{sec:charm_channel}

\begin{figure}[t]
\centering
\begin{lstlisting}[language=C++]
// C++ source file
// Chare init
void MyChare::init() {
	// Channel ID is 0
	channel = CkChannel(0, thisProxy[peer]);
}

// Data exchange
void MyChare::exchange() {
	channel.send(send_buf, size, CkCallbackResumeThread());
	channel.recv(recv_buf, size, CkCallbackResumeThread());
}
\end{lstlisting}
\caption{Example Usage of the Channel API in Charm++.}
\label{fig:charm_channel_code}
\end{figure}

The Channel API aims to improve communication performance
by avoiding the need of a host-side message in the GPU Messaging API.
A channel is first created between a pair of chare objects, with an ID
provided by the user that has to be unique in the program.
The channel can then be used to send and receive data by providing
the address of source or destination buffer, size of the data,
and a \texttt{CkCallback} object or a Charm++ future~\cite{charm_futures} that will be invoked
on completion of the channel primitive.
An example is shown in Figure~\ref{fig:charm_channel_code}, 
where a special type of callback, \texttt{CkCallbackResumeThread} is
used to suspend the calling thread to perform an asynchronous send or receive.
The thread will be awakened when the channel primitive completes, allowing
the chare to continue executing.
When a Charm++ future is provided, it can be later waited on for completion
similar to non-blocking MPI communication that uses an \text{MPI\_Request}.

\begin{figure}[t]
\centering
\includegraphics[width=\linewidth]{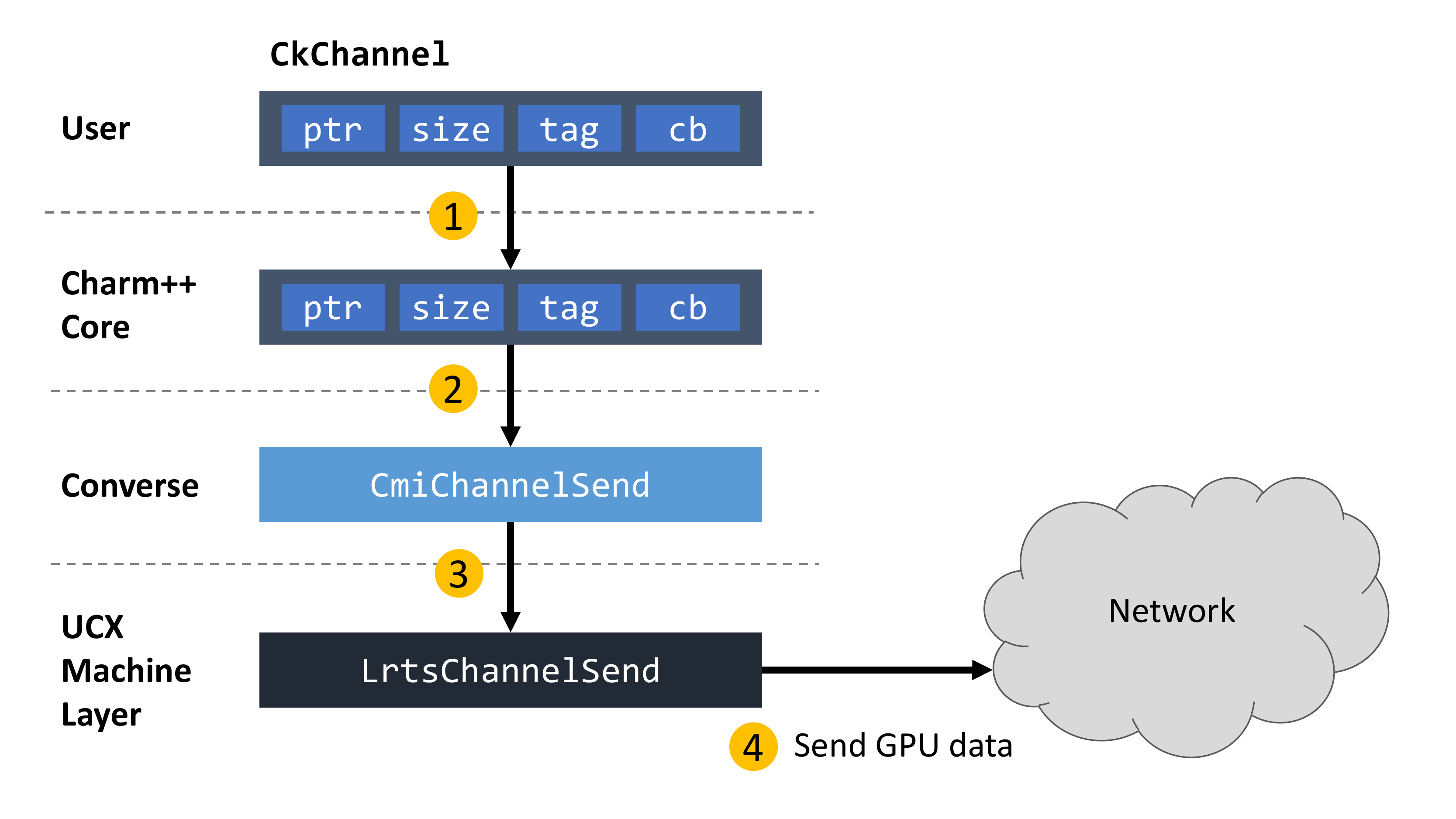}
\caption{Sender-side logic of the Channel API in Charm++.}
\label{fig:charm_channel_design}
\vspace{-10pt}
\end{figure}

The channel object maintains all information needed for GPU data exchanges between
the participating pair of chares, including the 64-bit tag.
Because both the sender and receiver chares keep track of which UCX tag is being
used for communication, and the destination GPU buffer address is provided by the
user in the channel receive call, there is no more need for a host-side message.
The receiver chare has all the information needed to post a receive for the incoming
GPU data, and the UCX machine layer is accessed almost directly by the Channel API
through \texttt{LrtsChannelSend} and \texttt{LrtsChannelRecv} calls.
These properties allow the Channel API to demonstrate better performance than the GPU Messaging API,
as discussed in Section~\ref{sec:perf}.
The sender side logic of the Channel API is illustrated in Figure~\ref{fig:charm_channel_design}.

\begin{figure}[t]
\centering
\includegraphics[width=\linewidth]{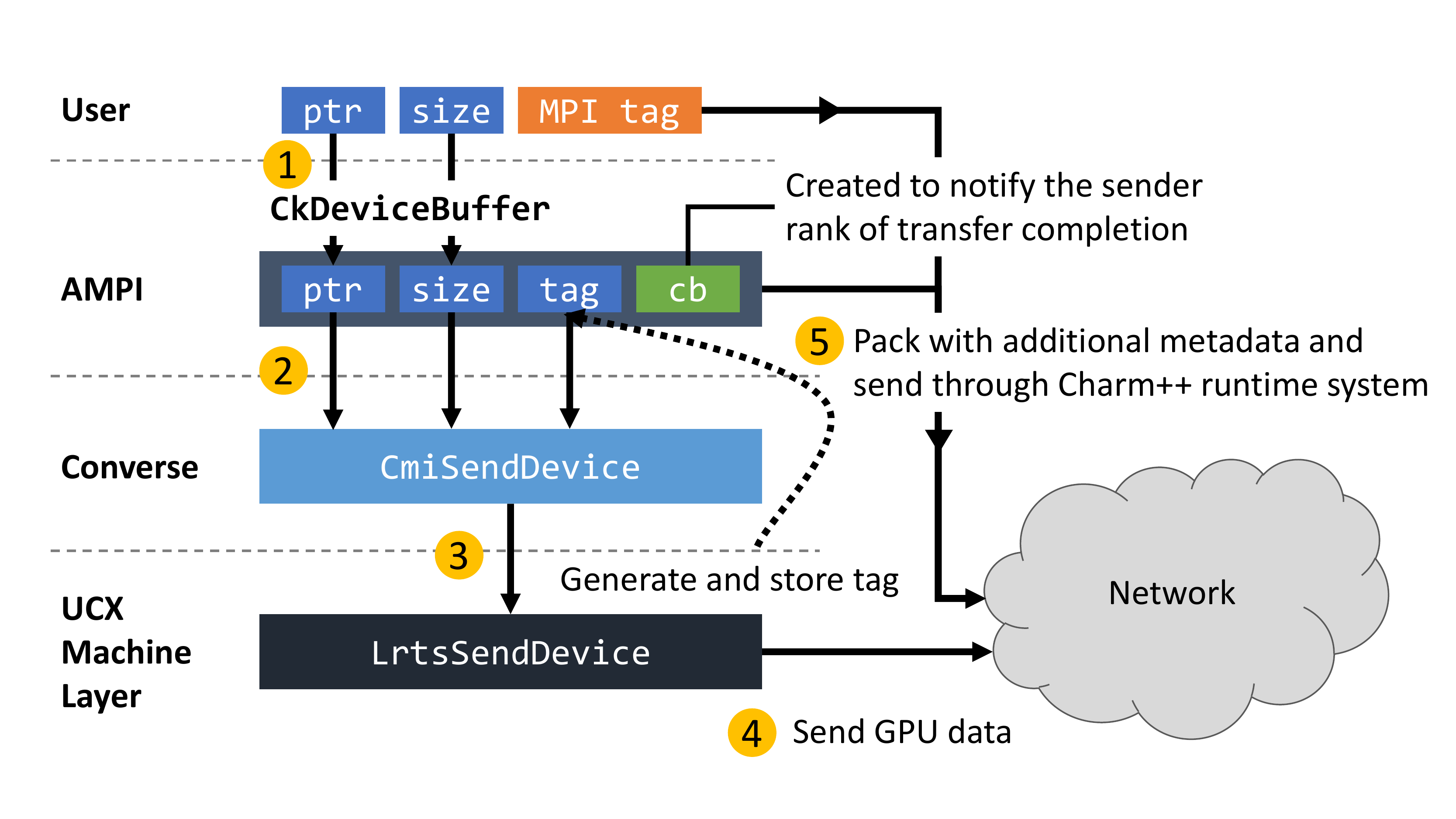}
\caption{Sender-side logic of GPU-aware communication in AMPI.}
\label{fig:ampi_gpu_design}
\vspace{-10pt}
\end{figure}

\subsection{Adaptive MPI}

Each AMPI rank is implemented as a chare object on top of the Charm++ runtime system,
to enable virtualization and adaptive runtime features such as load balancing.
Communication between AMPI ranks occurs through an exchange of AMPI messages
between the respective chare objects. An AMPI message
adds AMPI-specific data such as the MPI communicator and user-provided tag to
a Charm++ message. We modify how an AMPI message is created to integrate GPU-aware communication
with the GPU Messaging API and \texttt{CkDeviceBuffer} metadata object. This change is transparent
to the user, and GPU buffers can be directly provided to AMPI communication primitives
such as \texttt{MPI\_Send} and \texttt{MPI\_Recv} like any CUDA-aware MPI implementation.
We are also currently exploring the integration of the Channel API into AMPI to potentially further improve
performance, which can be done by creating a channel between the pair of chare objects
each mapped to an AMPI rank.

An AMPI application can send GPU data by invoking a MPI send call
with parameters including the address of the source buffer, number of elements and their datatype,
destination rank, tag, and MPI communicator. The chare object that manages
the destination rank is first determined, and the source buffer's address is checked
to see if it is located on GPU memory. A software cache containing addresses known to be on the GPU
is maintained on each PE to optimize this process.
Figure~\ref{fig:ampi_gpu_design} illustrates the mechanism that
is executed when the source buffer is found to be on the GPU,
where a \texttt{CkDeviceBuffer} object is first created in the AMPI runtime to store the
information provided by the user.
A Charm++ callback object is also created and stored as metadata, which is
used by AMPI to notify the sender rank when the communication is complete.
The source GPU buffer is sent in an identical manner as Charm++ through the UCX machine layer
with \texttt{LrtsDeviceSend}.
The tag that is needed by the receiver rank to post a receive for the incoming GPU data
is also generated and stored inside the \texttt{CkDeviceBuffer} object.
Note that this tag is separate from the MPI tag provided by the user,
which is used to match the host-side send and receive.

Because there are explicit receive calls in the MPI model in contrast to Charm++, there are two possible scenarios
regarding the host-side message that contains metadata: the message
arrives before the receive is posted, and vice versa.
If the message arrives first, it is stored in an unexpected
message queue, which is searched for a match when the receive is posted later.
If the receive is posted first, it is stored in a
request queue to be matched when the message arrives. The receive for the
incoming GPU data is posted after this match of the host-side message, with
\texttt{LrtsDeviceRecv} in the UCX machine layer.
Another Charm++ callback is created for the purpose of notifying the
destination rank, which is invoked by the machine layer when the GPU data arrives.

\begin{figure}[t]
\centering
\begin{lstlisting}[language=Python, morekeywords={}]
if not gpu_direct:
	# Host-staging mechanism (not GPU-aware)
	# Transfer GPU buffer to host memory and send
	charm.lib.CudaDtoH(h_send_buf, d_send_buf, size, stream)
	charm.lib.CudaStreamSynchronize(stream)
	channel.send(h_send_buf)

	# Receive on host and transfer to GPU
	h_recv_buf = channel.recv()
	charm.lib.CudaHtoD(d_recv_buf, h_recv_buf, size, stream)
	charm.lib.CudaStreamSynchronize(stream)
else:
	# GPU-aware communication
	# Send and receive GPU buffers directly
	channel.send(d_send_buf, size)
	channel.recv(d_recv_buf, size)
\end{lstlisting}
\caption{Channel-based communication in Charm4py. CUDA functions are included
in the Charm++ library as C++ functions and exposed through Charm4py's Cython layer.}
\label{fig:charm4py_gpu_code}
\vspace{-10pt}
\end{figure}

\subsection{Charm4py}\label{sec:design_charm4py}

GPU-aware communication is supported in Charm4py through its Channels feature,
which allows streamed communication between a pair of chares. This is what
the Channel API in Charm++ takes inspiration from, but GPU-aware communication
in Charm4py is currently built on top of the GPU Messaging API of Charm++,
not its Channel API. Like AMPI, the messaging mechanism in Charm4py is in the process
of being updated to be able to adopt Charm++'s Channel API.
While the Channels interface in Charm4py is in Python, its core
functionalities are implemented with Cython~\cite{cython} and the underlying
Charm++ runtime system is comprised of C++. Cython generates C extension modules
to support C constructs and types to be used with Python for interoperability and performance,
and is used extensively in the Charm4py runtime. The Cython layer is also
used to interface with the Charm++ runtime, which performs the bulk of the work
for GPU-aware communication with the UCX machine layer. Note that the Python
interface for UCX, UCX-Py~\cite{ucx_py}, is not used in this work as Charm4py can
directly utilize the UCX functionalities in C/C++ through the Charm++ runtime system.

Figure~\ref{fig:charm4py_gpu_code} compares our GPU-aware communication support
against the host-staging mechanism in a ping-pong exchange of GPU data.
Each of the two chares opens a channel to the other, which is used to
exchange data either on the host or GPU memory determined by the
\texttt{gpu\_direct} flag.
The host-staging version needs to explicitly move data between host and device memory
using the CUDA API, adding complexity to the programmer and degrading performance.
Note that the Charm4py channel send and receive calls are asynchronous; a send
call returns control as soon as it is initiated, and the coroutine posting
a receive is suspended and returns control back to the scheduler until the message arrives.
Such asynchronous mechanisms are implemented with futures~\cite{charm4py_futures}, a key component of Charm4py.

\begin{figure}[t]
\centering
\includegraphics[width=\linewidth]{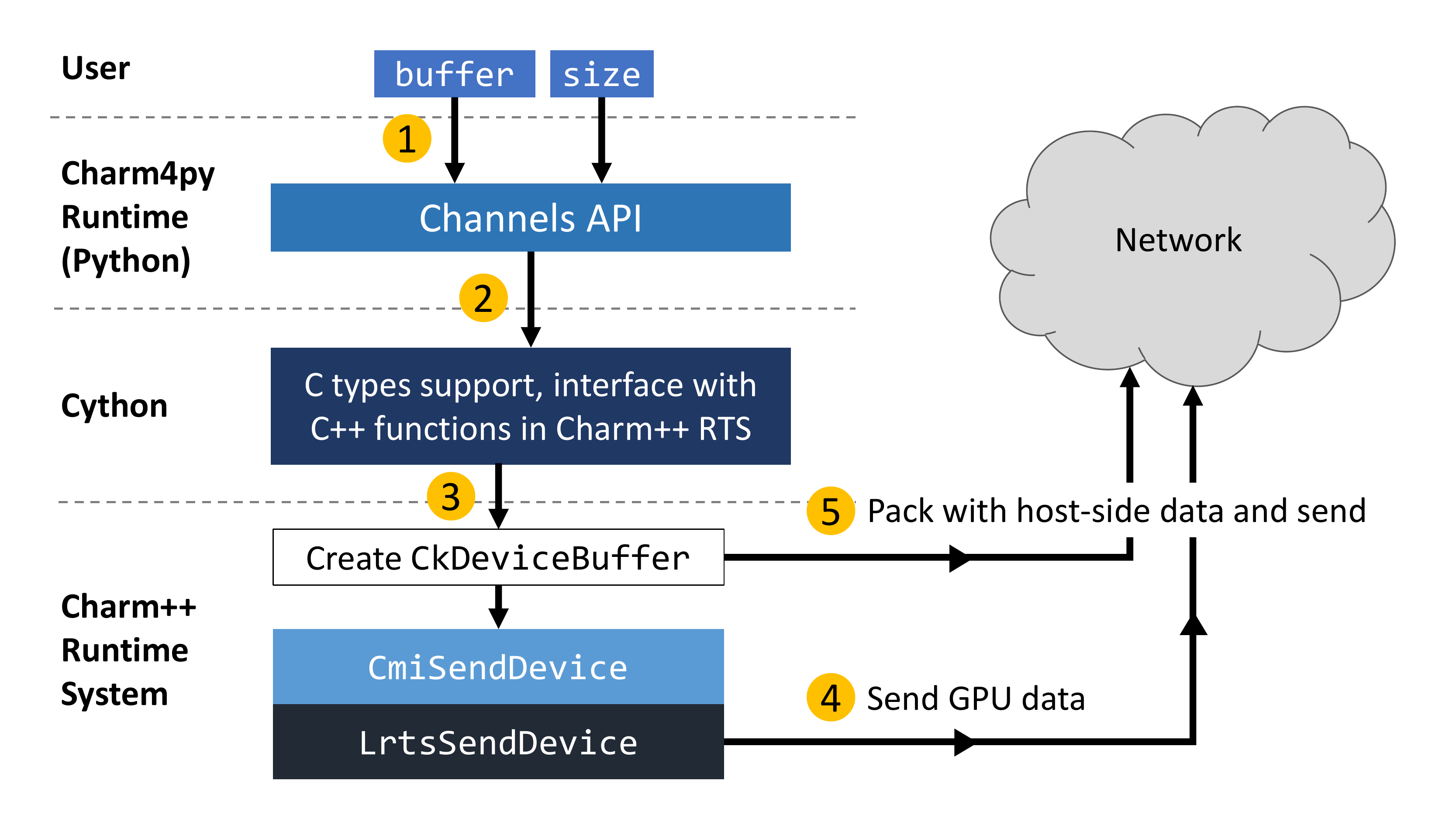}
\caption{Sender-side logic of GPU-aware communication in Charm4py.}
\label{fig:charm4py_gpu_design}
\vspace{-10pt}
\end{figure}

As can be seen from Figure~\ref{fig:charm4py_gpu_code}, addresses of the source and
destination GPU buffers can be directly provided to Charm4py's Channel API.
The address and size of the buffer are propagated to the Charm++ runtime
system through the Cython layer, which are used to construct the \texttt{CkDeviceBuffer}
metadata object. The steps after this point follow the GPU Messaging API,
where the UCX machine layer sends the source GPU buffer using the provided metadata.
The metadata is packed together with the host-side user data (if any) and Charm4py-specific
information and sent in a separate code path to the receiver object. This process is illustrated
in Figure~\ref{fig:charm4py_gpu_design}.

When the host-side message containing metadata about the GPU data transfer arrives,
\texttt{LrtsDeviceRecv} in the UCX machine layer posts a receive for the incoming GPU data.
A Charm++ callback is created and tied to the \texttt{LrtsDeviceRecv} function,
for the purpose of handling completion of the GPU communication.
The invocation of this callback fulfills the Charm4py future object that was created
on the channel receive call, which allows the user application (coroutine) to
continue executing.

\section{Performance Evaluation}\label{sec:perf}

In this section, we describe the hardware platform and software configurations,
as well as the set of micro-benchmarks and proxy application used to evaluate
the performance of our GPU-aware communication designs.

\begin{figure*}[t]
\centering
\subfloat[][Charm++]{\includegraphics[width=.33\linewidth]{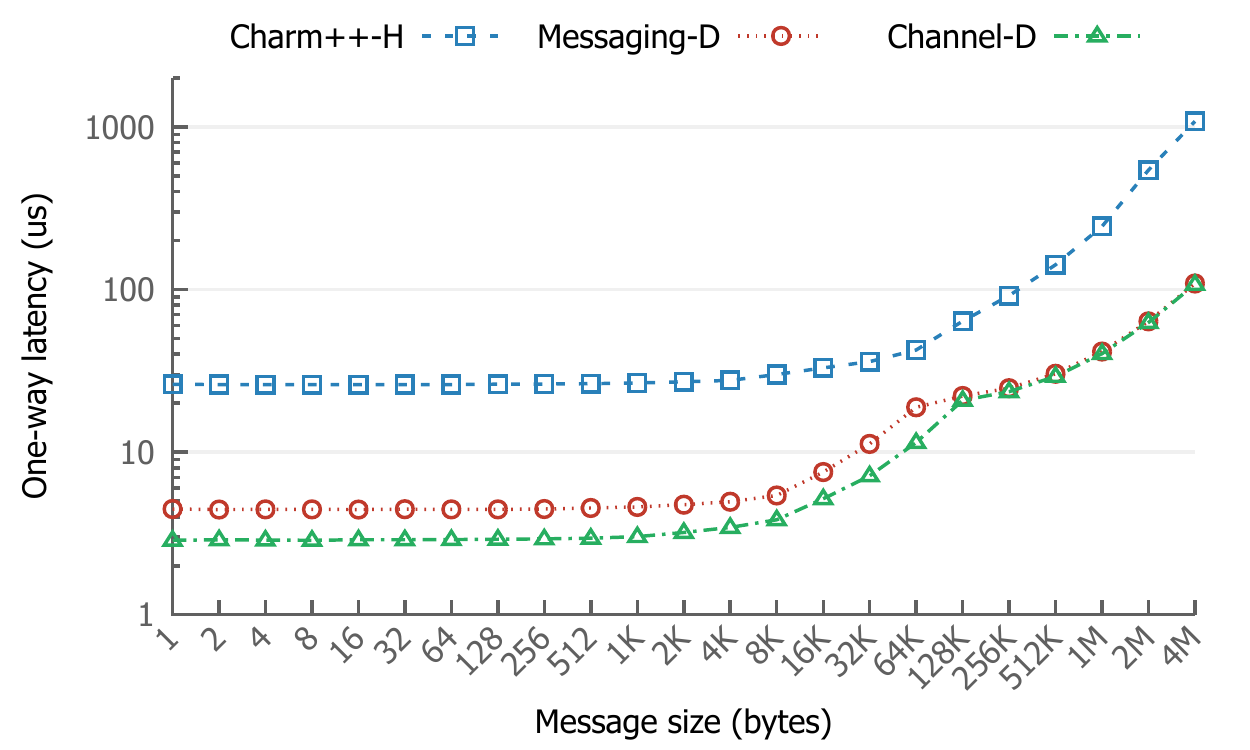}\label{fig:latency-charm-intra}}
\subfloat[][AMPI and OpenMPI]{\includegraphics[width=.33\linewidth]{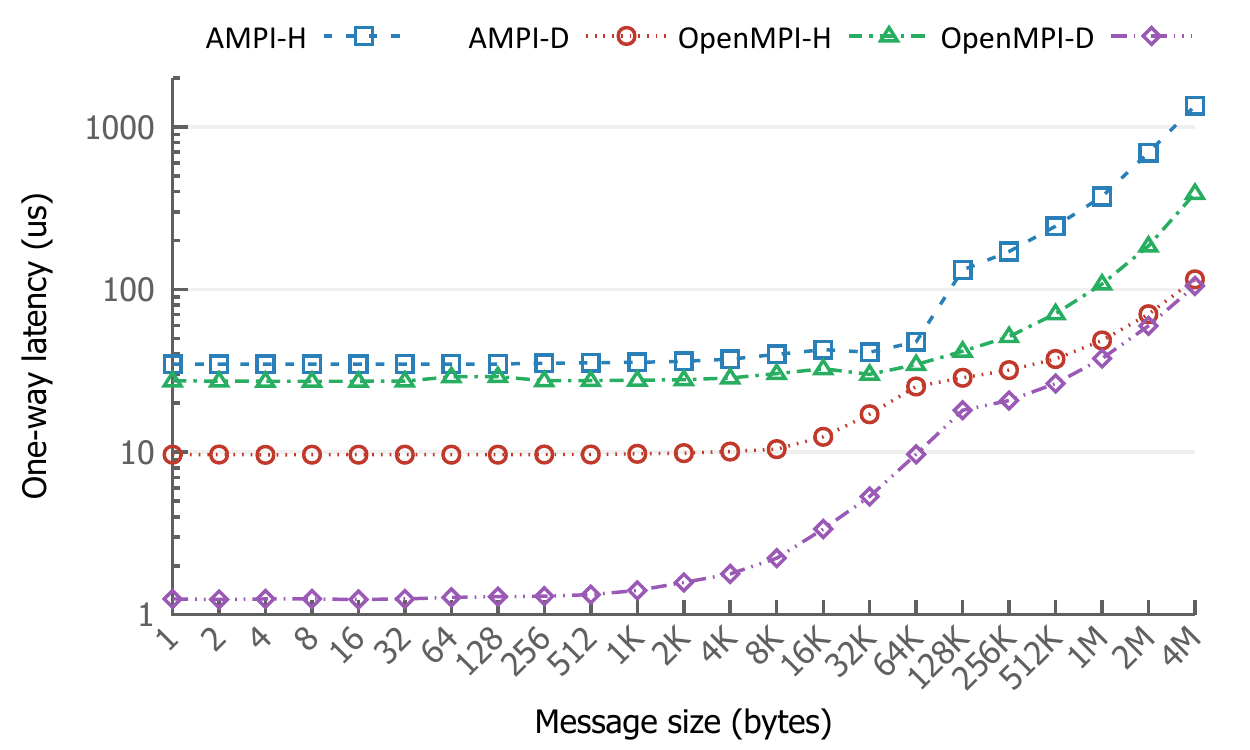}\label{fig:latency-mpi-intra}}
\subfloat[][Charm4py]{\includegraphics[width=.33\linewidth]{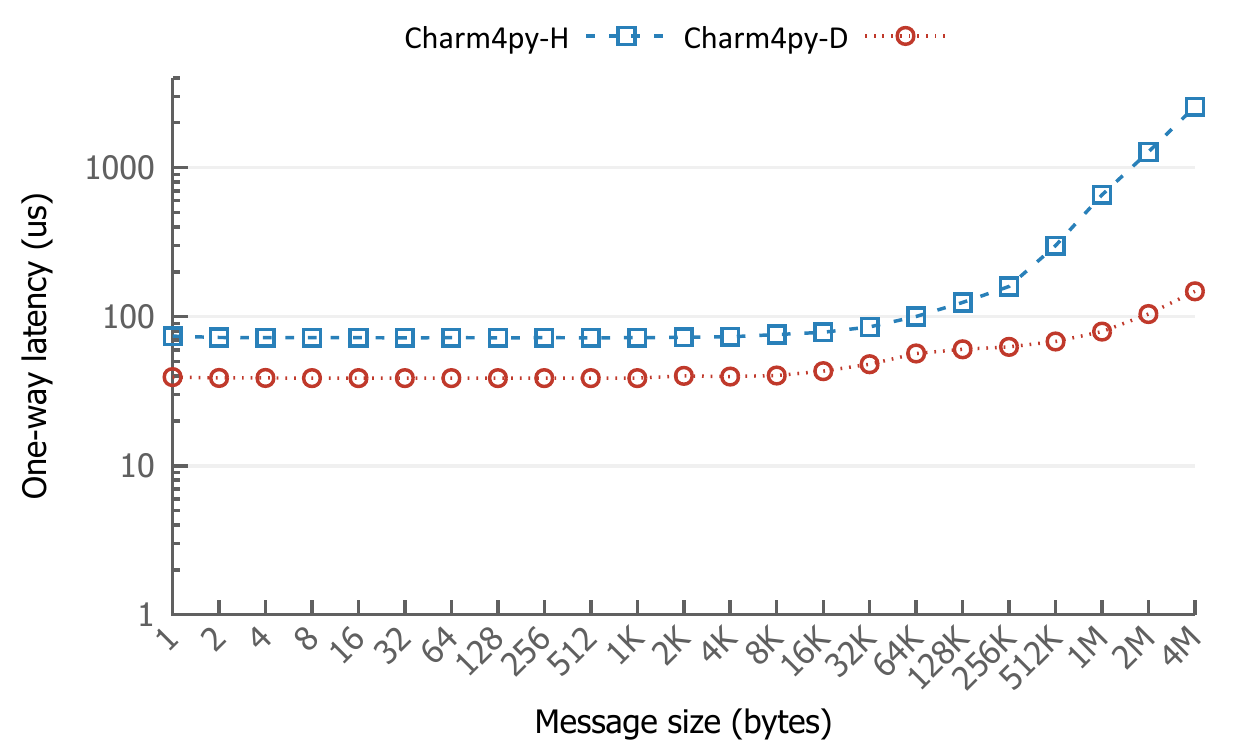}\label{fig:latency-charm4py-intra}}

\caption{Comparison of intra-node latency between host-staging and direct GPU-GPU mechanisms.}
\label{fig:latency-intra}
\vspace{-10pt}
\end{figure*}

\begin{figure*}[t]
\centering
\subfloat[][Charm++]{\includegraphics[width=.33\linewidth]{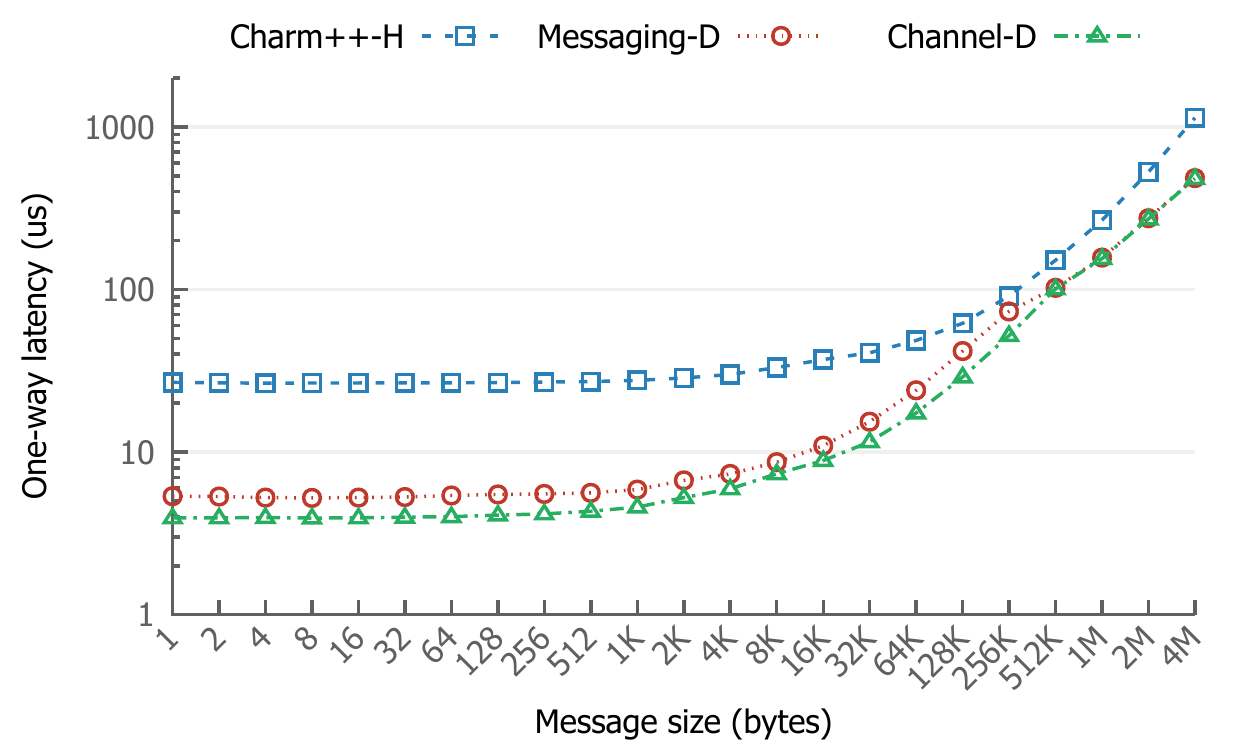}\label{fig:latency-charm-inter}}
\subfloat[][AMPI and OpenMPI]{\includegraphics[width=.33\linewidth]{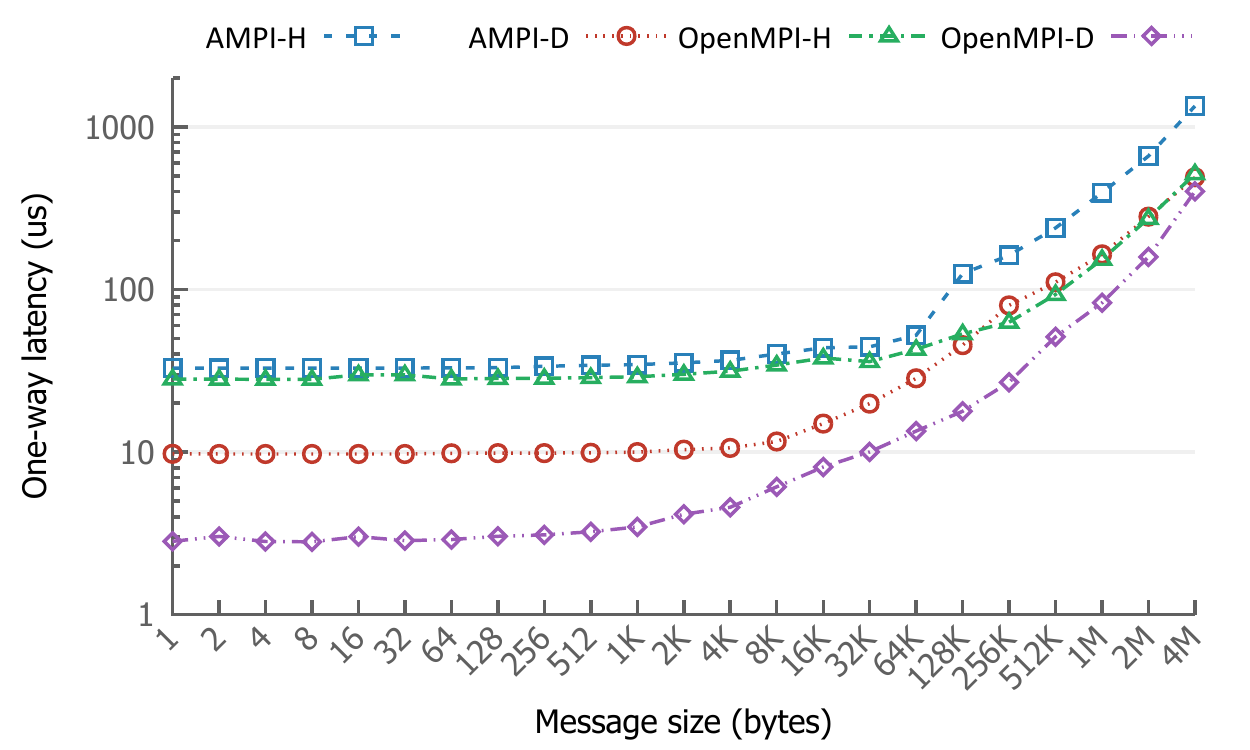}\label{fig:latency-mpi-inter}}
\subfloat[][Charm4py]{\includegraphics[width=.33\linewidth]{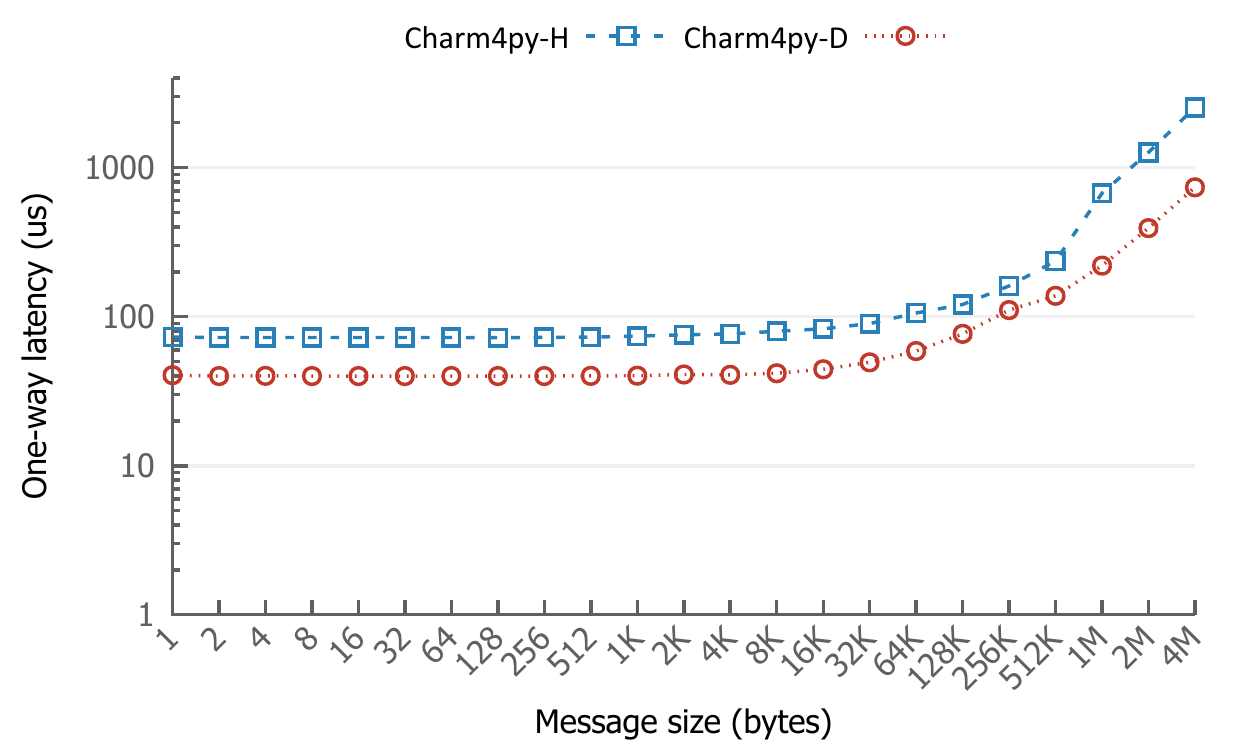}\label{fig:latency-charm4py-inter}}

\caption{Comparison of inter-node latency between host-staging and direct GPU-GPU mechanisms.}
\label{fig:latency-inter}
\vspace{-10pt}
\end{figure*}

\subsection{Experimental Setup}

The Summit supercomputer at Oak Ridge National Laboratory is used to
evaluate the performance of GPU-aware communication mechanisms
implemented in Charm++, AMPI and Charm4py. The experiments are scaled up to
256 nodes of Summit, where each IBM AC922 node contains
two IBM Power9 CPUs and six NVIDIA Tesla V100 GPUs. Each CPU is connected
to three GPUs, which are interconnected via NVLink with a theoretical peak
bandwidth of 50 GB/s. For a GPU to communicate with another GPU connected
to the other CPU, data needs to travel through the X-Bus that connects the CPUs
with a bandwidth of 64 GB/s. The network interconnect is based on Mellanox Enhanced
Data Rate (EDR) Infiniband, providing up to 12.5 GB/s of bandwidth.

Charm++, AMPI and Charm4py are configured to use the non-SMP build,
using one CPU core as the single PE for each process and one process
per GPU device. On a single node
of Summit, for example, up to six PEs and GPUs can be used.
To accurately evaluate the impact of GPU-awareness on communication performance
by separating communication from computation,
we do not employ overdecomposition and instead decompose the problem domain into
the same number of chare objects or AMPI ranks as the number of PEs and GPUs.

For reference, benchmark results with OpenMPI (which also maps one process to each GPU)
are provided along with the performance of AMPI.
Since both AMPI and OpenMPI utilize UCX to transfer GPU data, this comparison
isolates the performance differential incurred by the layers above UCX.
We expect the performance of AMPI
to be slower because of the overheads associated with message-driven execution
such as copies between the user application and the runtime system.
We expect the performance of AMPI to be slower than OpenMPI because of the overheads
associated with message-driven execution in the Charm++ RTS such as memory copies between the user application
and the underlying runtime. This is in contrast to OpenMPI which can directly
utilize UCX for communication.



\subsection{Micro-benchmarks}

To evaluate the performance of point-to-point communication primitives involving GPU memory,
we adapt the widely used OSU micro-benchmark suite~\cite{eurompi12-omb-gpu} to Charm++
and Charm4py.
For Charm++, we compare the performance of both the GPU Messaging API and
the Channel API to the host-staging method. We expect the Channel API to perform
better due to the lack of a metadata message delaying the receive for the GPU buffer.
We also compare our GPU-aware communication mechanisms against host-staging
in AMPI and OpenMPI.
Performance results are presented with both axes in log-scale,
comparing the GPU-aware version(s) of the benchmark (suffixed with D) against
the host-staging version (suffixed with H). Results with the two different mechanisms in Charm++,
the GPU Messaging API and Channel API, are provided as Messaging-D and Channel-D, respectively.

\begin{figure*}[t]
\centering
\subfloat[][Charm++]{\includegraphics[width=.33\linewidth]{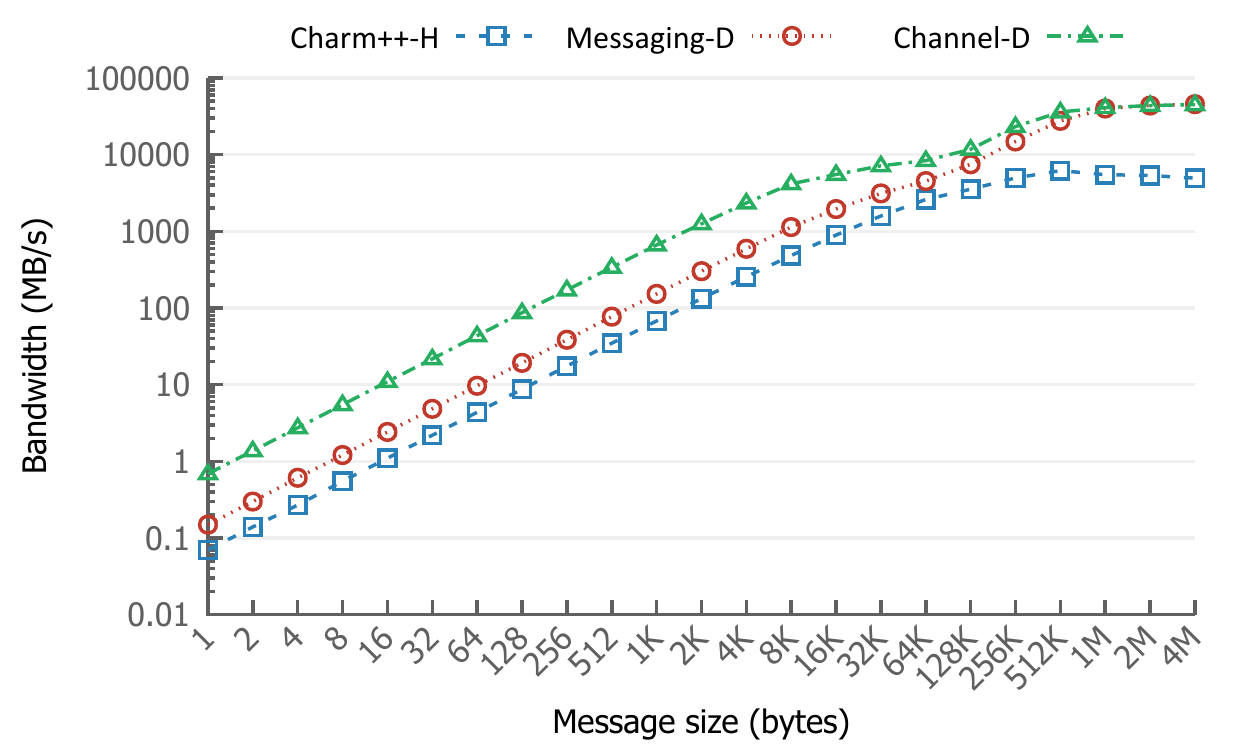}\label{fig:bandwidth-charm-intra}}
\subfloat[][AMPI and OpenMPI]{\includegraphics[width=.33\linewidth]{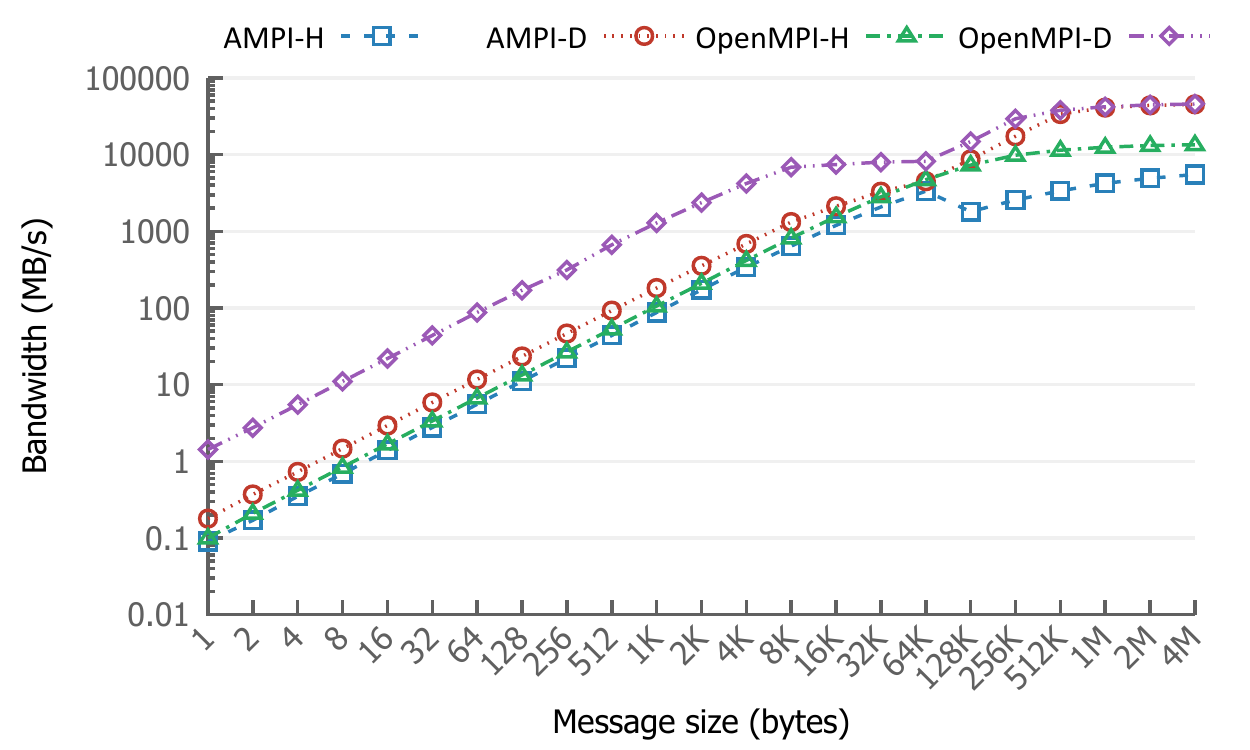}\label{fig:bandwidth-mpi-intra}}
\subfloat[][Charm4py]{\includegraphics[width=.33\linewidth]{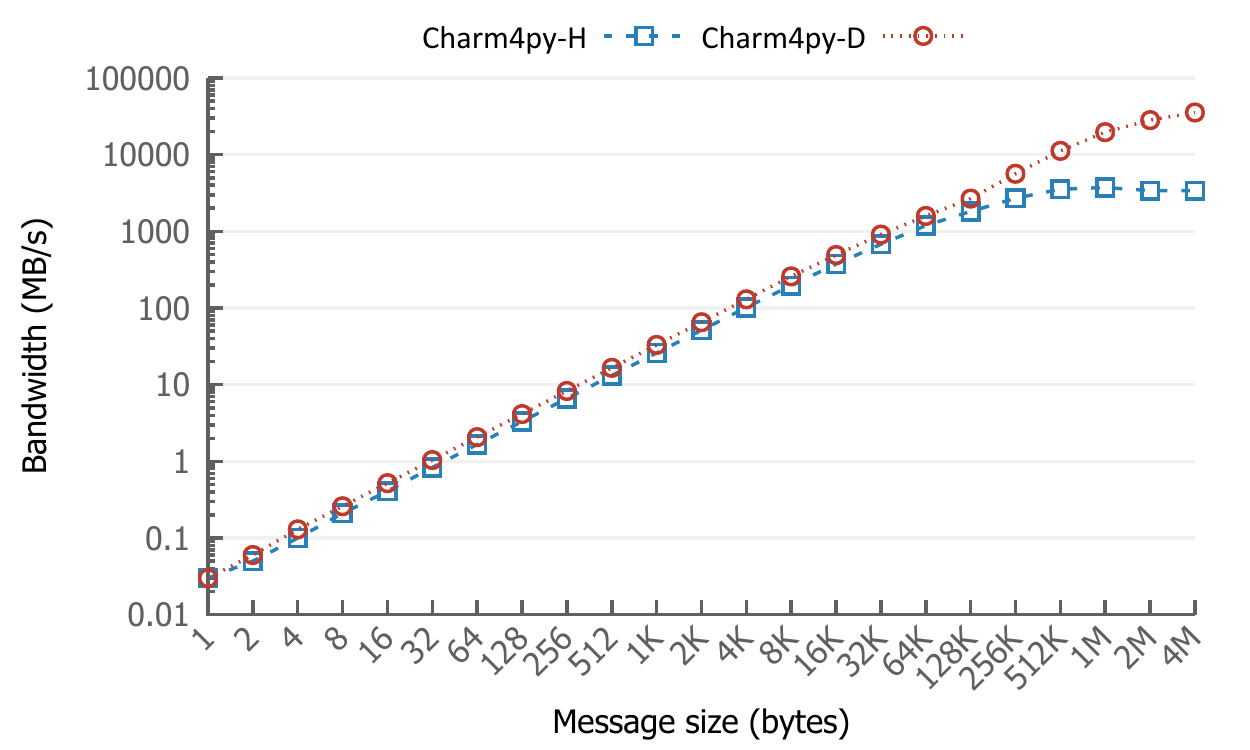}\label{fig:bandwidth-charm4py-intra}}

\caption{Comparison of intra-node bandwidth between host-staging and direct GPU-GPU mechanisms.}
\label{fig:bandwidth-intra}
\vspace{-10pt}
\end{figure*}

\begin{figure*}[t]
\centering
\subfloat[][Charm++]{\includegraphics[width=.33\linewidth]{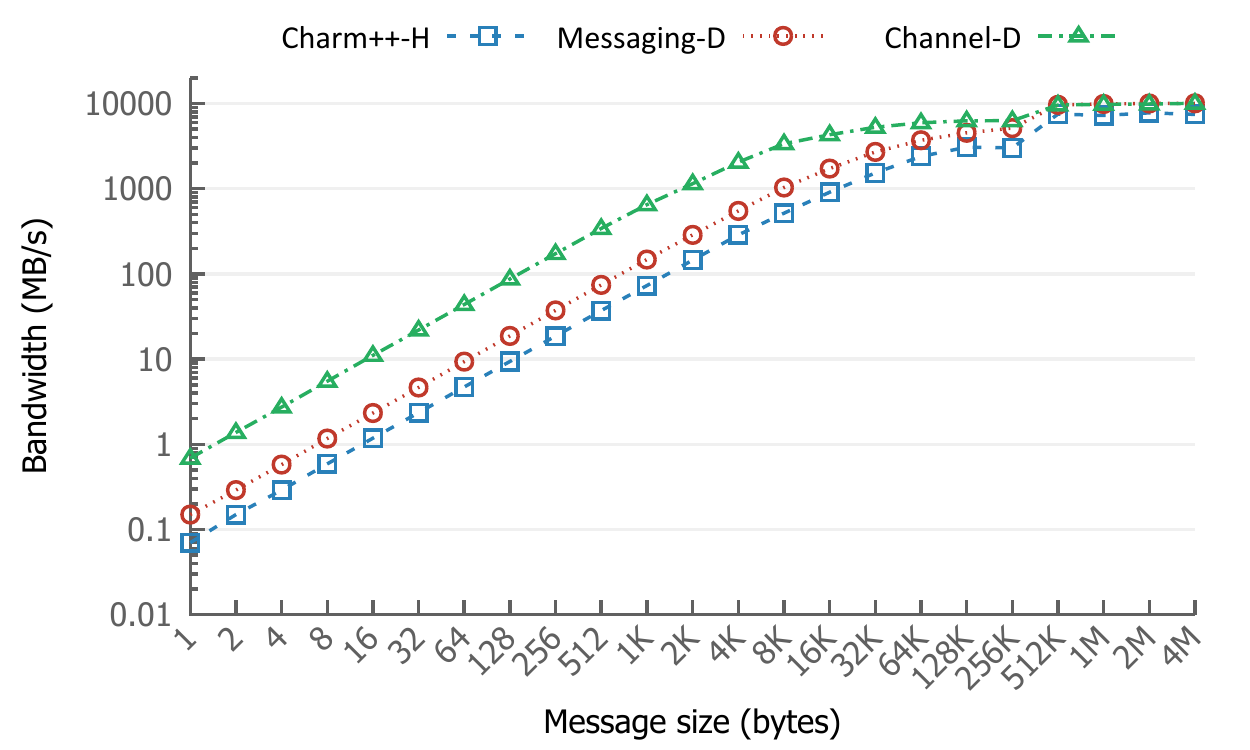}\label{fig:bandwidth-charm-inter}}
\subfloat[][AMPI and OpenMPI]{\includegraphics[width=.33\linewidth]{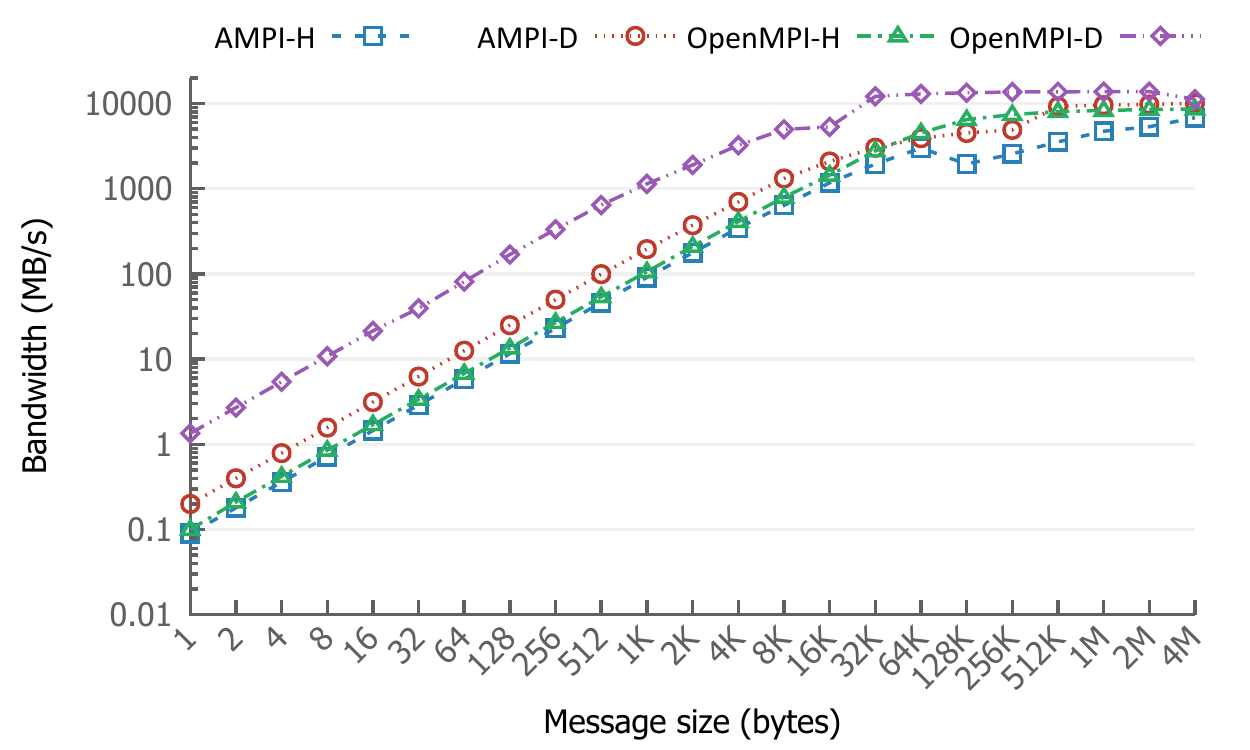}\label{fig:bandwidth-mpi-inter}}
\subfloat[][Charm4py]{\includegraphics[width=.33\linewidth]{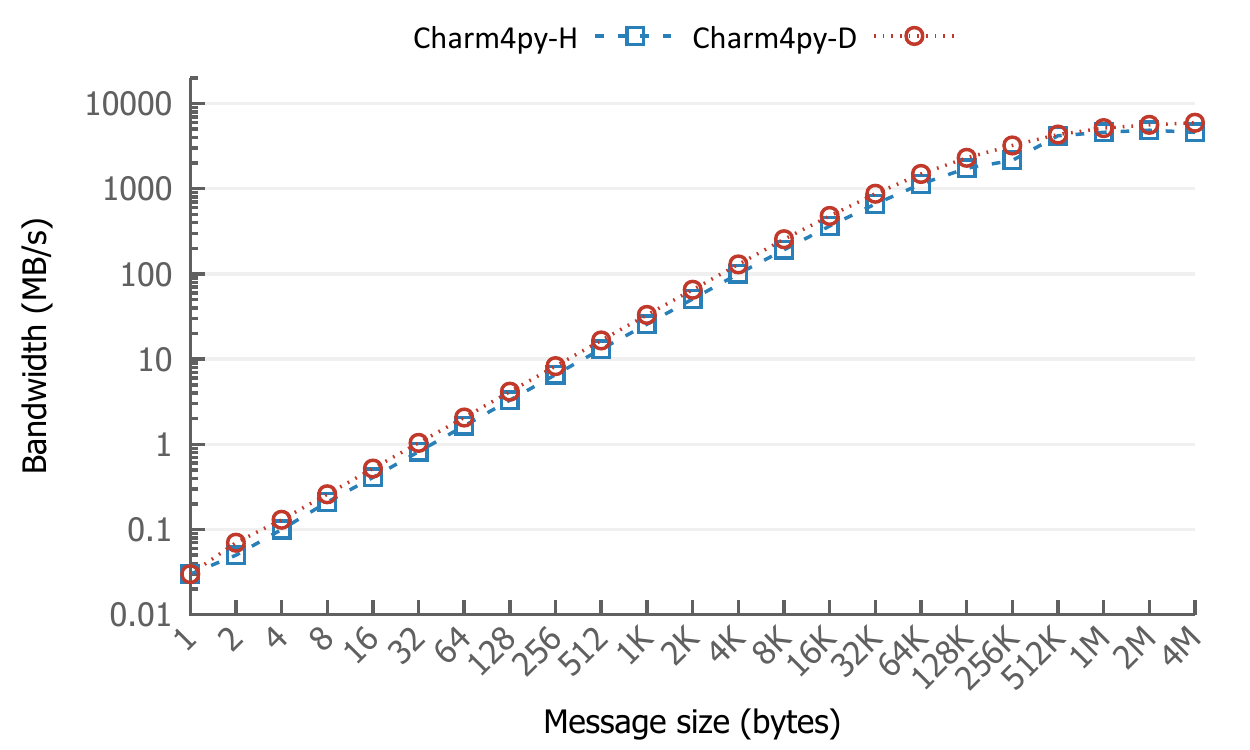}\label{fig:bandwidth-charm4py-inter}}

\caption{Comparison of inter-node bandwidth between host-staging and direct GPU-GPU mechanisms.}
\label{fig:bandwidth-inter}
\end{figure*}

\begin{table*}[!t]
\small
\renewcommand{\arraystretch}{1}
\caption{Improvement in latency and bandwidth with GPU-aware communication.}
\begin{center}
\begin{tabular}{lccccccc} \toprule[0.1em]
\multirow{2}{*}{\textbf{Improvement}} & \multirow{2}{*}{\textbf{Type}} & \multicolumn{3}{c}{\textbf{Intra-node}} & \multicolumn{3}{c}{\textbf{Inter-node}}\\
& & \textbf{Charm++} & \textbf{AMPI} & \textbf{Charm4py} & \textbf{Charm++} & \textbf{AMPI} & \textbf{Charm4py}\\
\midrule[0.05em]
\multirow{2}{*}{\textbf{Latency}} & Range & 3.1x {\textendash} 10.1x & 1.9x {\textendash} 11.7x & 1.8x {\textendash} 17.4x & 1.5x {\textendash} 6.8x & 1.8x {\textendash} 3.5x & 1.5x {\textendash} 3.4x\\
& Eager & 9.1x & 3.6x & 1.9x & 6.8x & 3.4x & 1.8x\\
\midrule[0.05em]
\textbf{Bandwidth} & Range & 1.7x {\textendash} 10.1x & 1.3x {\textendash} 10.0x & 1.3x {\textendash} 10.5x & 1.3x {\textendash} 9.4x & 1.3x {\textendash} 2.6x & 1.0x {\textendash} 1.5x\\
\bottomrule[0.1em]
\end{tabular}
\label{tab:microbenchmark_speedup}
\end{center}
\vspace{-10pt}
\end{table*}

\subsubsection{Latency}

The OSU latency benchmark repeats ping-pong iterations for different message sizes,
where the sender sends a message to the receiver and waits for a reply.
Once the message arrives, the receiver sends a message with the same size
back to the sender, completing the round trip. GPU-aware communication allows
the message buffers to be supplied directly to the communication primitives,
whereas the host-staging version requires additional data transfers
between the host and device.

Figures~\ref{fig:latency-intra} and \ref{fig:latency-inter} illustrate the improvements
in intra-node and inter-node latency with GPU-awareness in Charm++, AMPI and Charm4py.
The range of performance improvements in the latency benchmark is summarized in Table~\ref{tab:microbenchmark_speedup},
where the achieved speedups with small messages using the eager protocol are denoted in a separate row. As the Channel API performs better than the GPU Messaging API,
its results are used for comparison against the host-staging mechanism.
The observed improvement in latency increases with message size
with large messages in all three programming models, as the host-staging mechanism
suffers from performance degradation caused by host memory copies performed by the Charm++ runtime system.

Although the performance of AMPI improves substantially with GPU-aware communication,
it does not quite match the latency of CUDA-aware OpenMPI. To further investigate
this issue, we isolate the time taken in UCX by taking advantage of the modular
structure of the UCX machine layer. We can easily disable the \texttt{CmiSend/RecvDevice}
calls in the Converse layer and directly invoke the receive handlers, mimicking instant completion of
the respective communication routines. This allows us to determine
the time that is taken outside of UCX, which turns out to be about 8~$\mu s$. This tells us
that the GPU data transfer itself with UCX has a latency of less than 2~$\mu s$,
similar to OpenMPI. Thus most of the overhead is AMPI-specific, which includes multiple factors:
message packing and unpacking, additional host-side message which contains metadata, Charm++
callback invocations, and the fact that the receiver rank cannot post a receive until
the metadata message is received. There are also a couple of heap memory allocations that
are needed to store metadata in the UCX machine layer to enable asynchronous communication.
We plan to further analyze and optimize the code
to get AMPI's performance as close to OpenMPI as possible.
Redesigning the AMPI code path to utilize the newly developed Channel API
of Charm++ instead of the GPU Messaging API could also bring AMPI's performance closer to OpenMPI.

It should be noted that the detection of
the GDRCopy library by UCX is essential in order to achieve low latencies with small
messages, which is not included in the default library search path on Summit.
With the rendezvous protocol, UCX switches to the CUDA IPC transport for intra-node
transfers, and to the pipelined host-staging mechanism that stages GPU data on host
memory in chunks for inter-node communication.

\begin{figure*}[!t]
\centering
\subfloat[][Weak scaling, overall time]{\includegraphics[width=.25\linewidth]{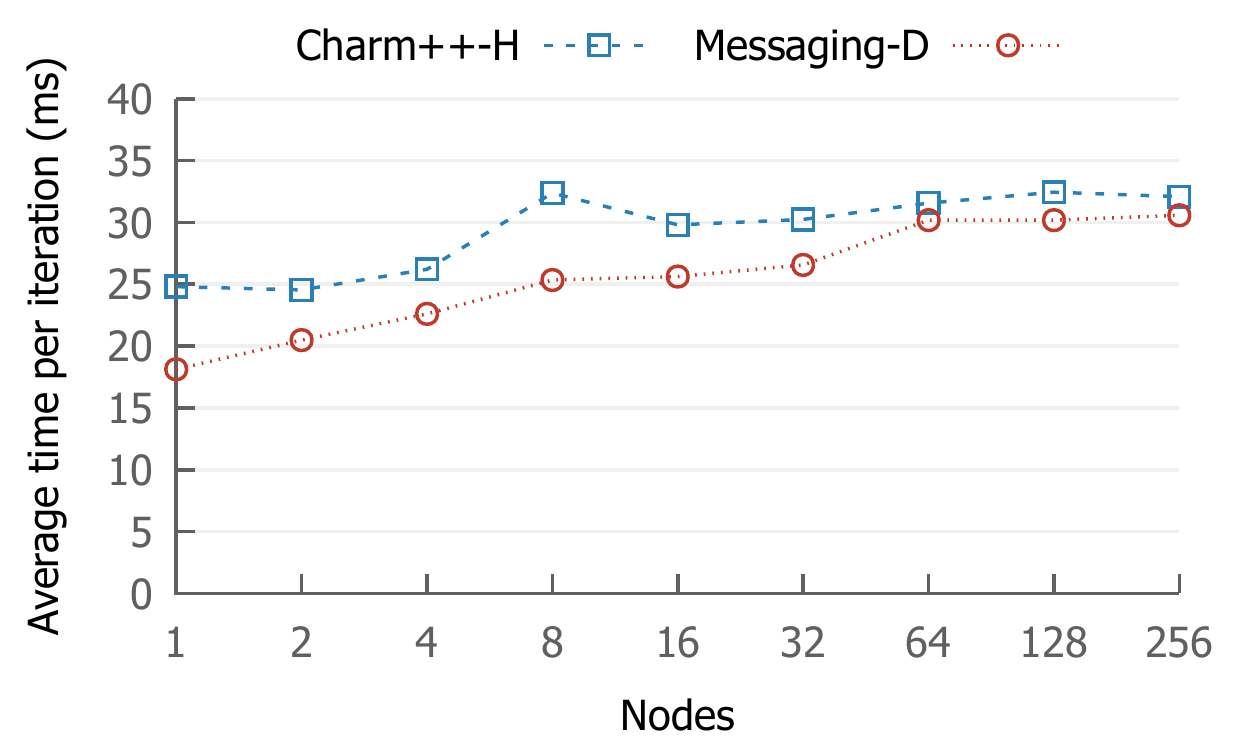}\label{fig:jacobi3d-charm-weak-total}}
\subfloat[][Weak scaling, comm. time]{\includegraphics[width=.25\linewidth]{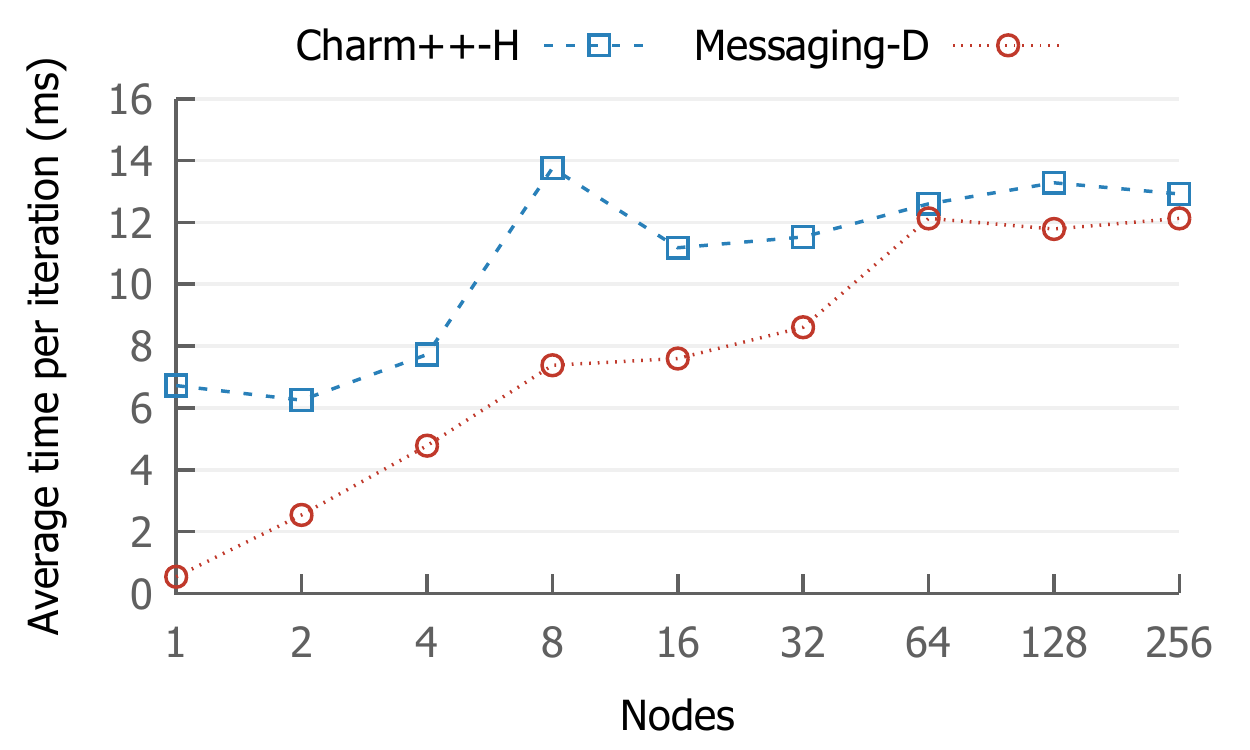}\label{fig:jacobi3d-charm-weak-comm}}
\subfloat[][Strong scaling, overall time]{\includegraphics[width=.25\linewidth]{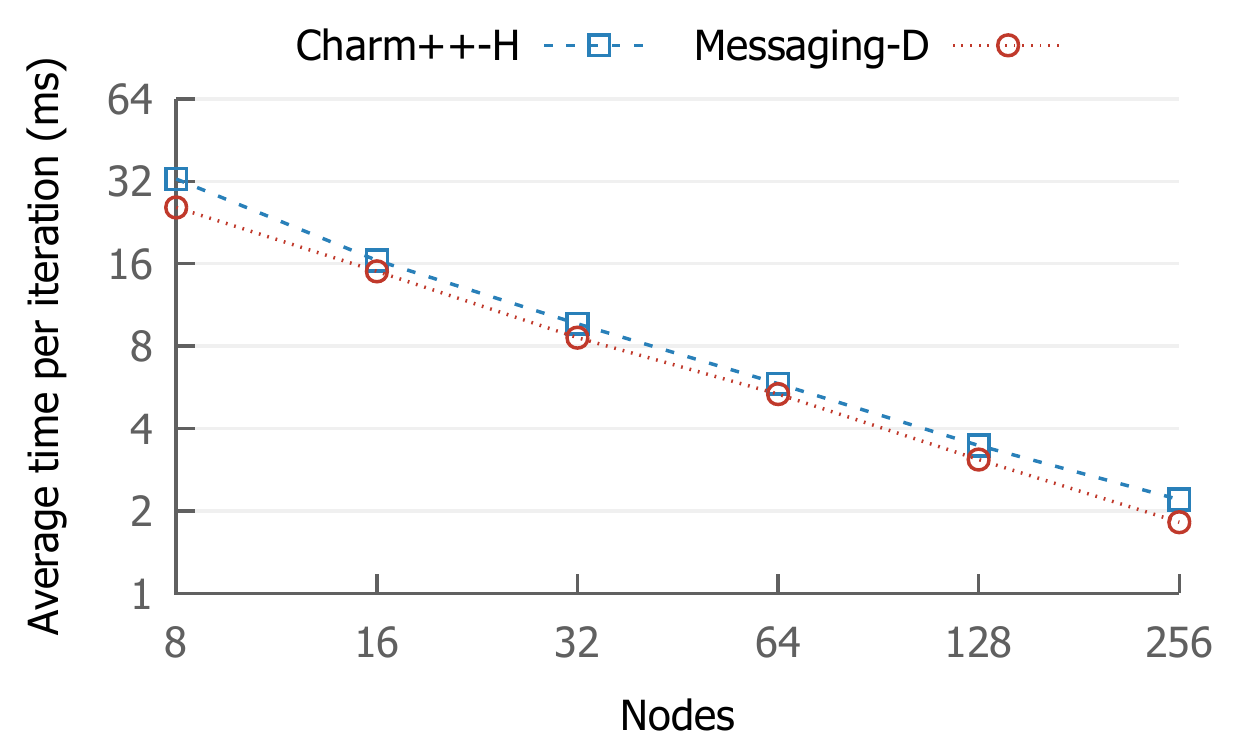}\label{fig:jacobi3d-charm-strong-total}}
\subfloat[][Strong scaling, comm. time]{\includegraphics[width=.25\linewidth]{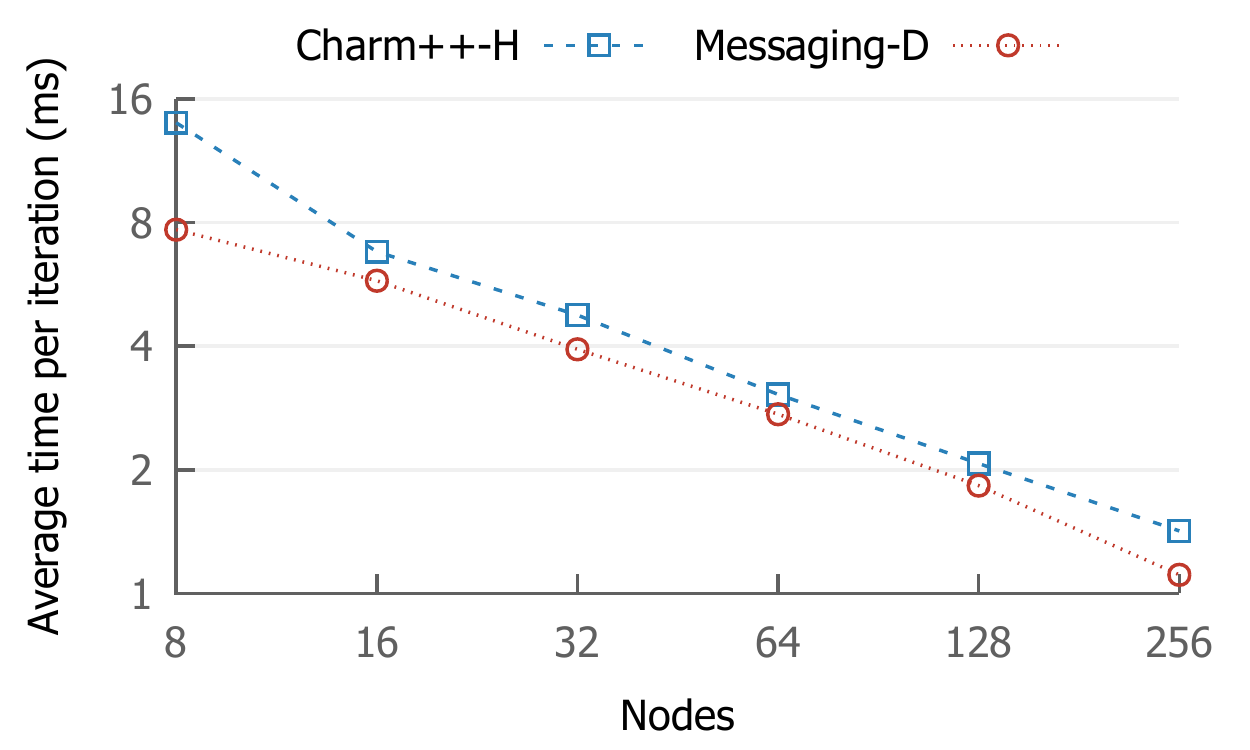}\label{fig:jacobi3d-charm-strong-comm}}

\caption{Comparison of Charm++ Jacobi3D performance between host-staging and direct GPU-GPU mechanisms.}
\label{fig:jacobi3d-charm}
\vspace{-10pt}
\end{figure*}

\begin{figure*}[!t]
\centering
\subfloat[][Weak scaling, overall time]{\includegraphics[width=.25\linewidth]{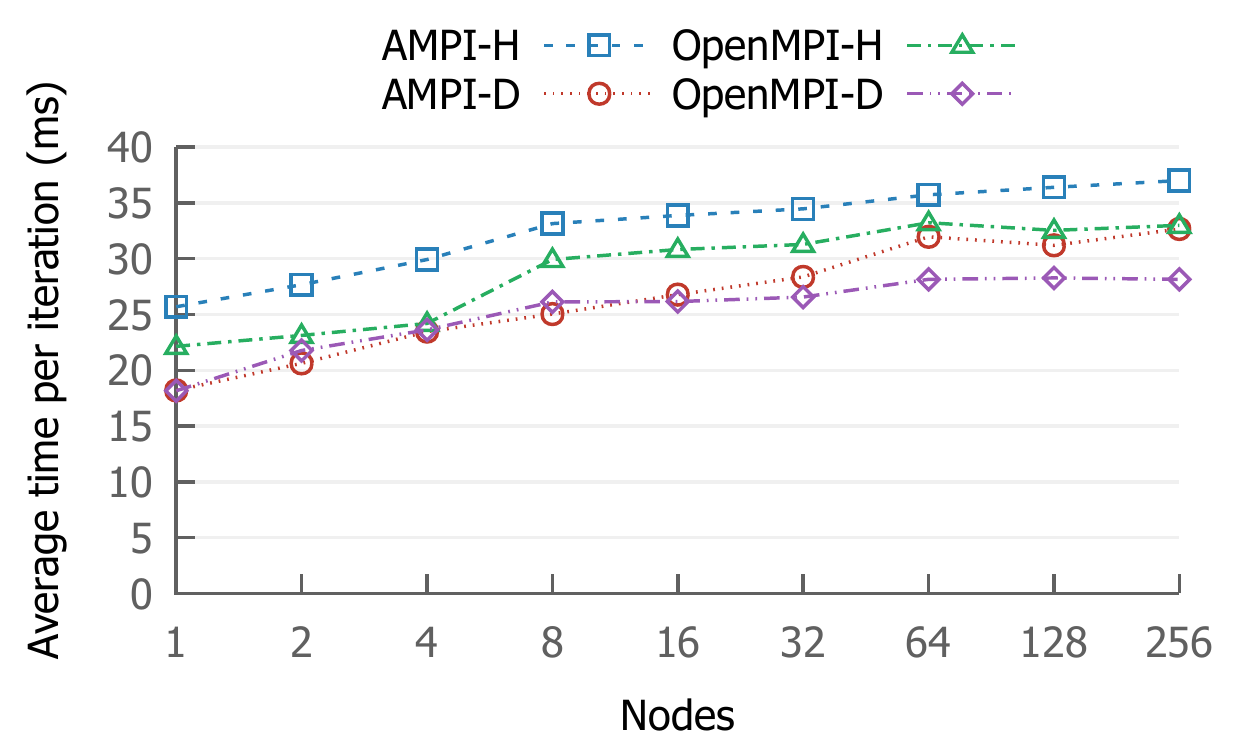}\label{fig:jacobi3d-mpi-weak-total}}
\subfloat[][Weak scaling, comm. time]{\includegraphics[width=.25\linewidth]{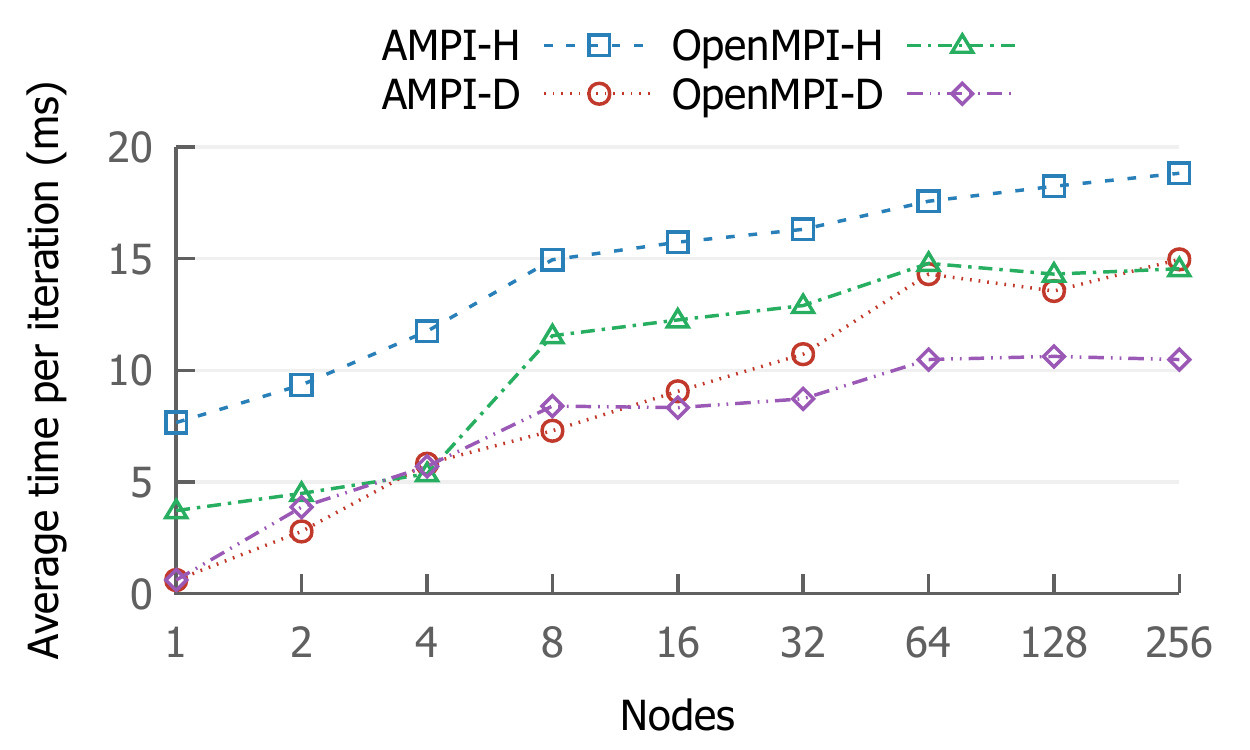}\label{fig:jacobi3d-mpi-weak-comm}}
\subfloat[][Strong scaling, overall time]{\includegraphics[width=.25\linewidth]{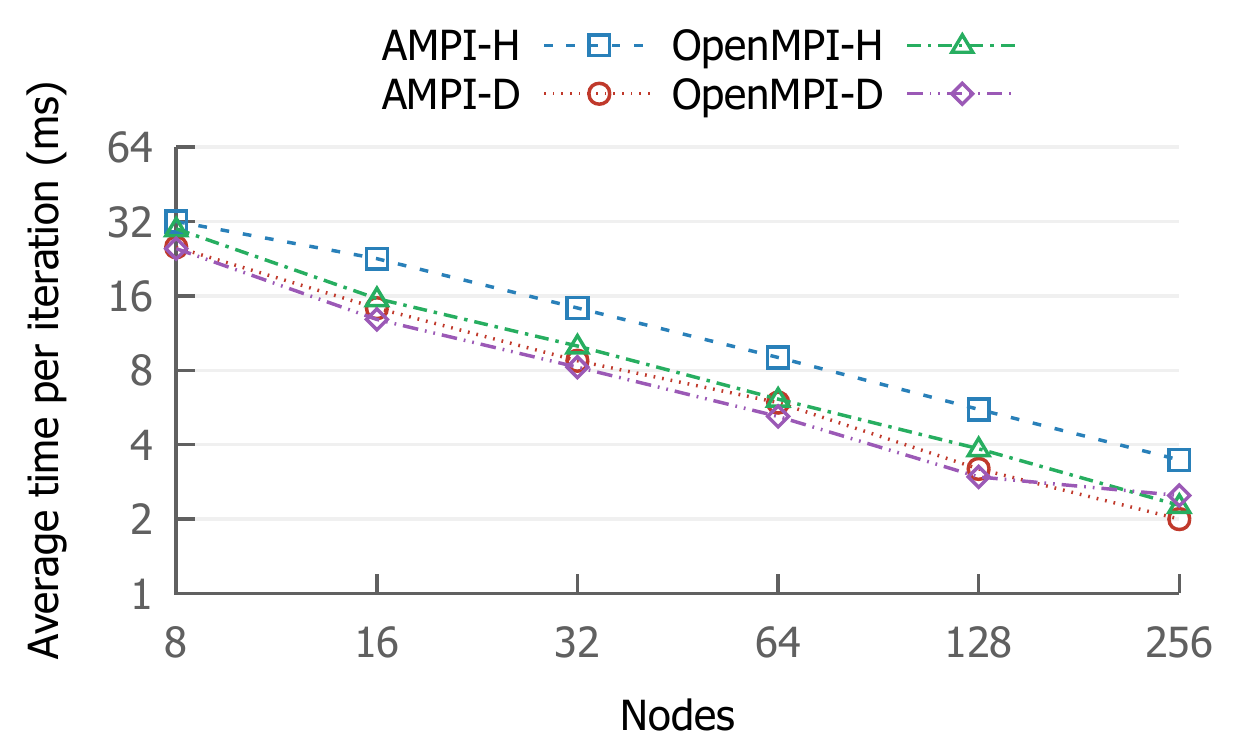}\label{fig:jacobi3d-mpi-strong-total}}
\subfloat[][Strong scaling, comm. time]{\includegraphics[width=.25\linewidth]{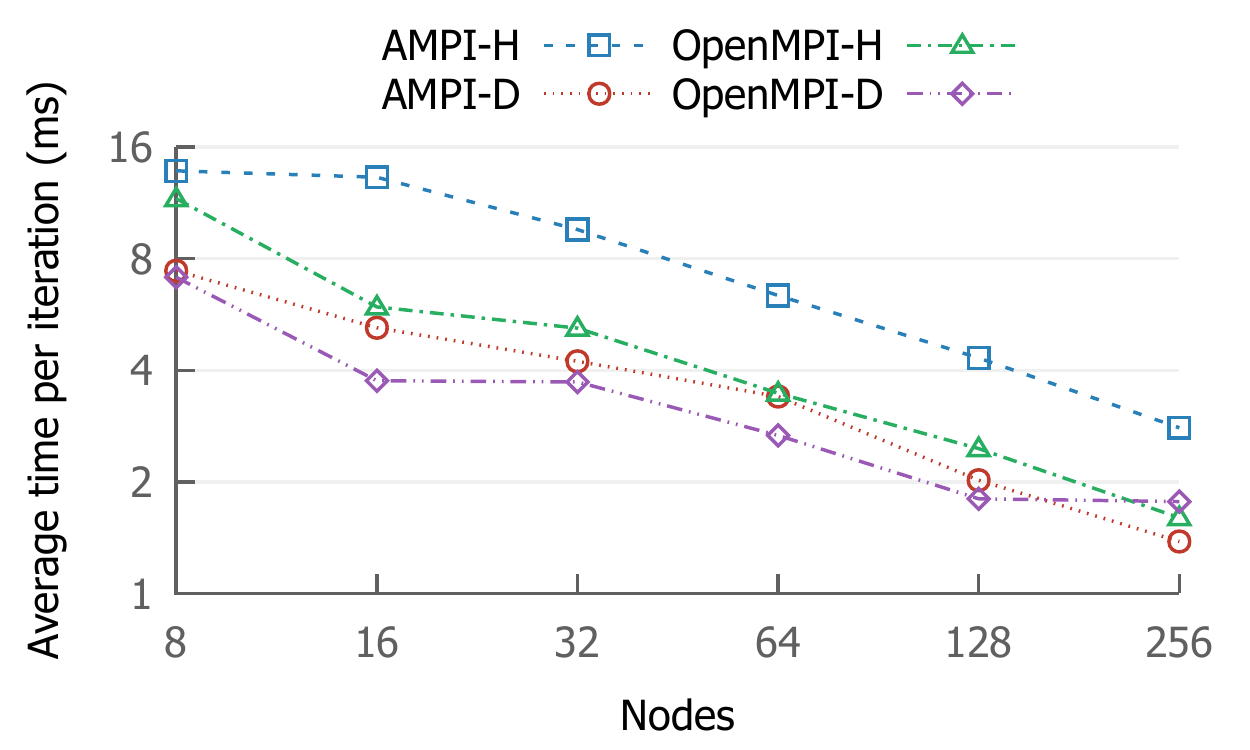}\label{fig:jacobi3d-mpi-strong-comm}}

\caption{Comparison of AMPI Jacobi3D performance between host-staging and direct GPU-GPU mechanisms.}
\label{fig:jacobi3d-mpi}
\vspace{-10pt}
\end{figure*}

\begin{figure*}[!t]
\centering
\subfloat[][Weak scaling, overall time]{\includegraphics[width=.25\linewidth]{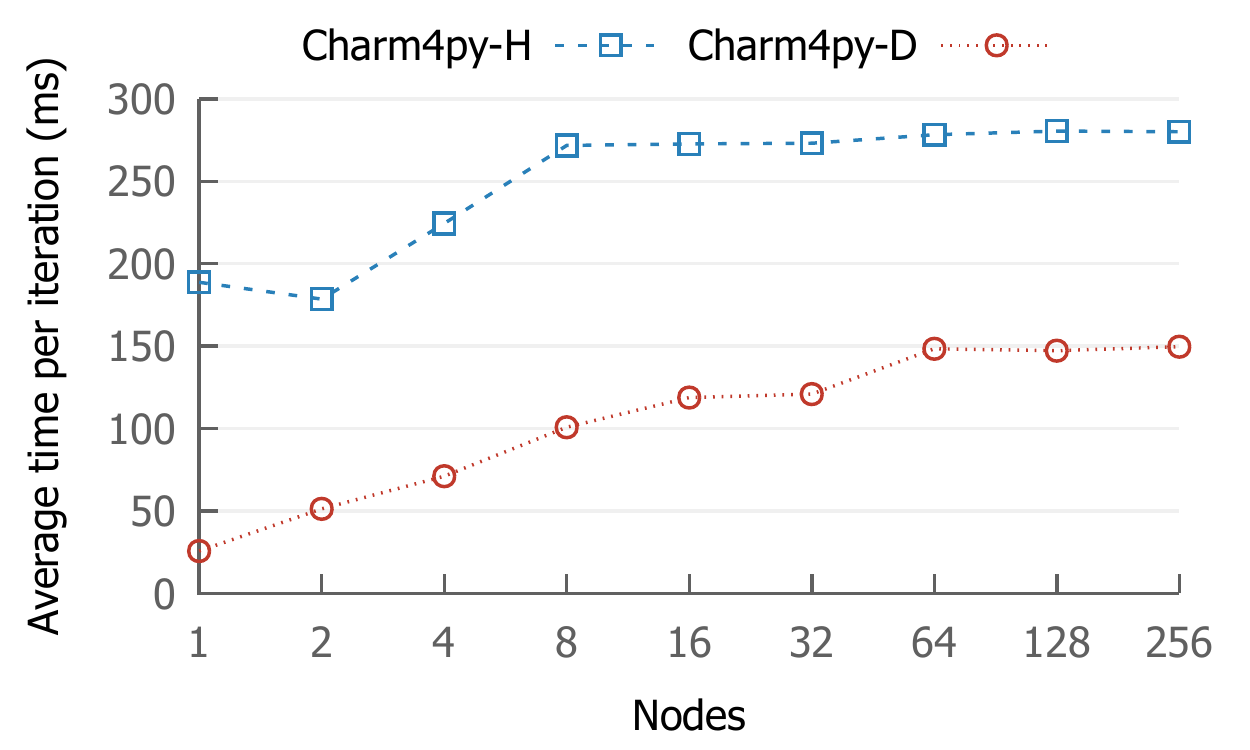}\label{fig:jacobi3d-charm4py-weak-total}}
\subfloat[][Weak scaling, comm. time]{\includegraphics[width=.25\linewidth]{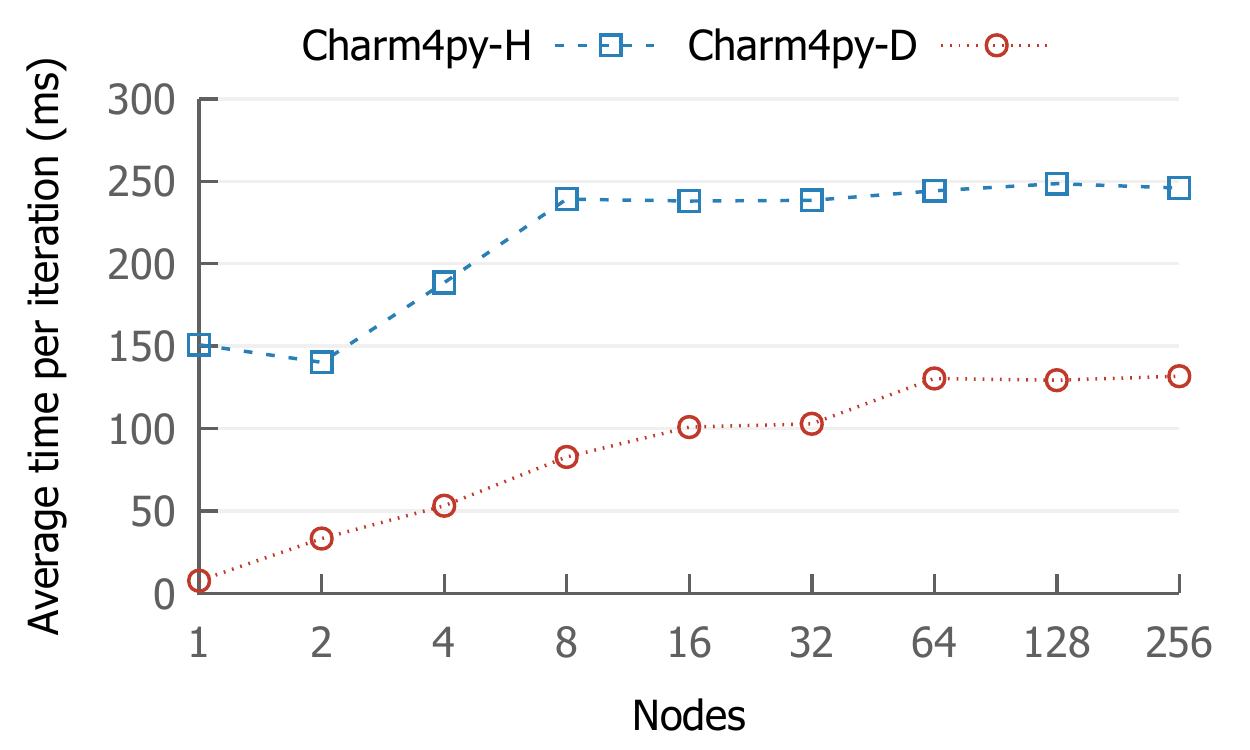}\label{fig:jacobi3d-charm4py-weak-comm}}
\subfloat[][Strong scaling, overall time]{\includegraphics[width=.25\linewidth]{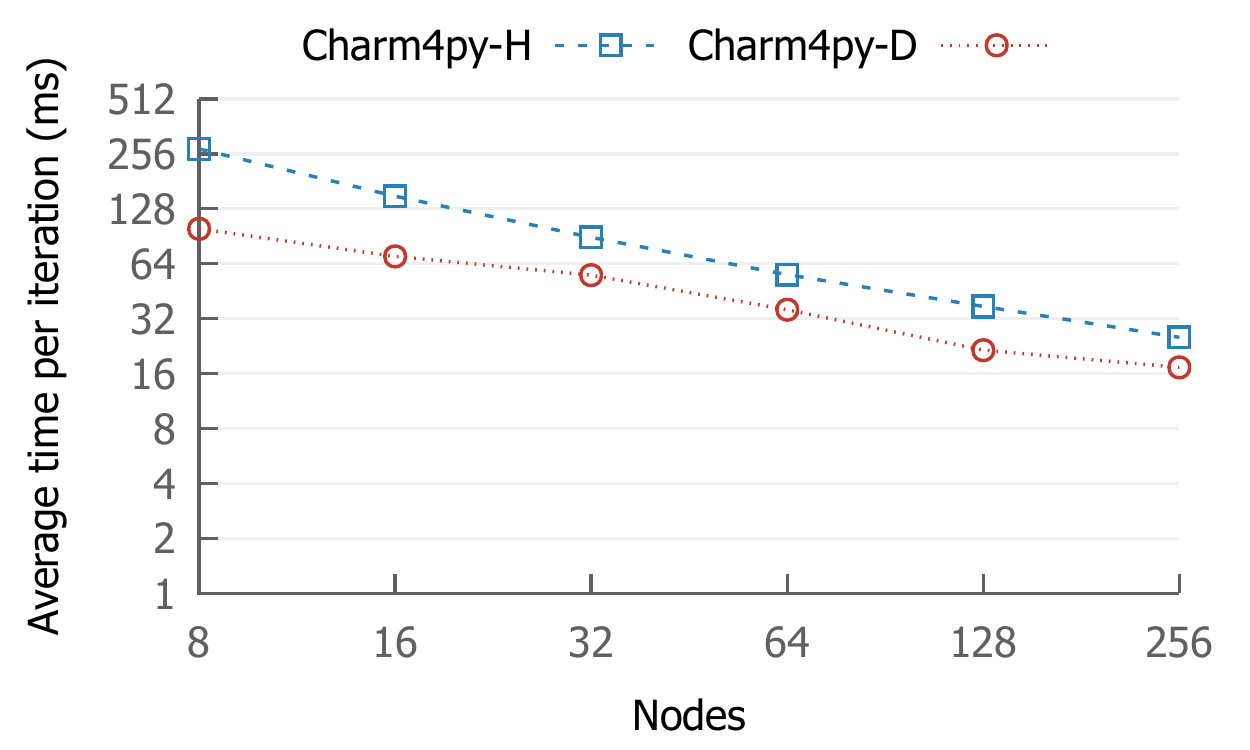}\label{fig:jacobi3d-charm4py-strong-total}}
\subfloat[][Strong scaling, comm. time]{\includegraphics[width=.25\linewidth]{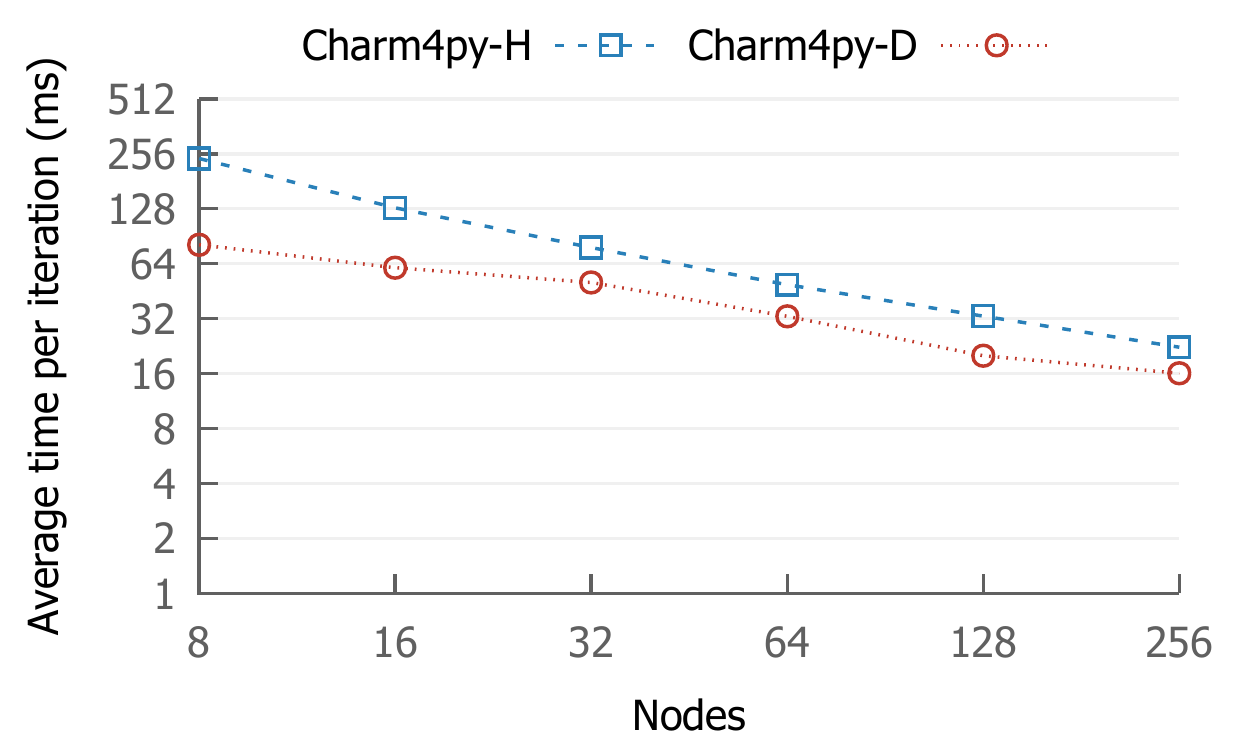}\label{fig:jacobi3d-charm4py-strong-comm}}

\caption{Comparison of Charm4py Jacobi3D performance between host-staging and direct GPU-GPU mechanisms.}
\label{fig:jacobi3d-charm4py}
\vspace{-10pt}
\end{figure*}

\subsubsection{Bandwidth}

In the OSU bandwidth benchmark, the sender performs a number of back-to-back non-blocking sends specified by
the window size for each message size, then waits for a reply from the receiver. The receiver
performs the reverse, posting multiple non-blocking receives followed by a send.
The increases in bandwidth achieved by our GPU-aware communication mechanisms are illustrated in
Figures~\ref{fig:bandwidth-intra} and \ref{fig:bandwidth-inter}, with the range of improvement
summarized in Table~\ref{tab:microbenchmark_speedup}.

Charm++ and AMPI achieve close to the maximum attainable bandwidth
(50~GB/s for intra-node, 12.5~GB/s for inter-node), with Charm++ demonstrating up to 44.7~GB/s and 10~GB/s,
and AMPI up to 45.4~GB/s and 10~GB/s for intra-node and inter-node, respectively.
As expected in Charm++, the Channel API (Channel-D) achieves higher bandwidth compared to the GPU Messaging API (Messaging-D)
especially with smaller messages.
This is because the Channel API is able to post the receive for the incoming GPU buffer without being delayed by the metadata message.
It is worth noting that the host-staging version of AMPI (AMPI-H) suffers a degradation
in bandwidth at 128~KB due to a sudden drop in performance, which is an issue that is being investigated.
Charm4py's bandwidth only reaches 35.5~GB/s for intra-node and 6.0~GB/s for inter-node in the given range of message sizes,
but we observe that it keeps increasing as messages become larger than 4~MB.

\subsection{Proxy Application: Jacobi3D}

To assess the impact of GPU-aware communication on application performance,
we implement a proxy application, Jacobi3D, on all three parallel programming
models: Charm++, AMPI, and Charm4py. Jacobi3D performs the Jacobi iterative
method in a three-dimensional space, using CUDA kernels to perform
stencil computations on the GPU. The problem domain is decomposed into
equal-size cuboid blocks, using a decomposition strategy that minimizes surface area to reduce communication volume.
Each block exchanges its halo data on the GPU with up to six neighbors, which are either provided
directly to the communication primitives (if GPU-aware) or staged through
host memory. Note that Jacobi3D is configured to
run for a set number of iterations without convergence checks, to be able to evaluate
the performance of point-to-point communication.

We evaluate both weak and strong scaling performance of Jacobi3D using up to
256 nodes (1,536 GPUs) of Summit, comparing the per-iteration execution times and communication times
of the host-staging and GPU-aware communication mechanisms.
Jacobi3D is weak-scaled with a base domain size of $1,536^3$ double values and
each dimension doubled in x, y, z order. Strong scaling experiments are performed
from eight to 256 nodes while maintaining a constant domain size of $3,072^3$ doubles.

\subsubsection{Charm++}

Figure~\ref{fig:jacobi3d-charm} shows the weak and strong scaling performance
of the Charm++ versions of Jacobi3D. With weak scaling, the implementation with
the GPU Messaging API (Messaging-D) demonstrates a speedup between 1.1x and 12.4x
in communication performance, with the largest speedup obtained on a single
node. This is an expected result as the improvements in latency and bandwidth
are more pronounced for intra-node communication. The improved communication
performance entails reductions in the overall execution time, ranging between
5\% and 37\%. The relative speedup obtained with GPU-aware communication decreases
as the number of nodes increases, as slower inter-node communication starts to dominate
intra-node communication. With strong scaling, the improvement in communication
performance ranges between 12\% and 82\% and overall iteration time between
9\% and 27\%, with the largest speedup obtained on a single node.

\subsubsection{AMPI}

Figure~\ref{fig:jacobi3d-mpi} illustrates the weak and strong scaling performance of
the AMPI versions of Jacobi3D, with the performance of OpenMPI provided as reference.
With weak scaling, GPU-awareness improves the communication performance by factors
between 1.3x and 12.8x, accelerating the overall performance up to 41\%.
The GPU-aware communication performance in AMPI is similar to that of OpenMPI up to 16 nodes,
but starts to fall behind at larger scales. We suspect that this is due to the
additional metadata exchange performed in AMPI, whose performance impact becomes more pronounced
at large node counts.
With strong scaling, AMPI achieves a speedup between 1.9x and 2.6x in communication
performance and an improvement in overall iteration time between 27\% and 74\%.

\subsubsection{Charm4py}

The weak and strong scaling performance of Charm4py are depicted in Figure~\ref{fig:jacobi3d-charm4py}.
As the support for GPU-aware communication in Charm4py significantly improves performance especially
for large messages as seen in Figures~\ref{fig:latency-charm4py-intra} and \ref{fig:latency-charm4py-inter},
communication performance is improved by factors between between 1.9x and 19.7x with
weak scaling. Because communication performance has a greater impact on the overall performance in
Charm4py compared to other parallel programming models, we observe speedups in overall execution time
between 1.9x and 7.3x. With strong scaling, the improvement in communication performance
ranges between 1.4x and 3.0x, resulting in speedups between 1.5x and 2.7x
in the overall iteration times.

\section{Related Work}\label{sec:related}

There have been many publications on supporting GPU-aware communication in
the context of parallel programming models.
Works from the MVAPICH group~\cite{jcs11-mvapich2-gpu, ipdpsw12-cuda_ipc, icpp13-potluri} utilize
CUDA and GPUDirect technologies to optimize inter-GPU communication in MPI.
Hanford et al.~\cite{exampi20-hanford} highlights shortcomings of current
GPU communication benchmarks and shares experiences with tuning different MPI implementations.
Khorassani et al.~\cite{mpi_openpower} evaluates the performance of various
MPI implementations on GPU-accelerated OpenPOWER systems.
Chen et al.~\cite{upc} proposes compiler extensions to support
GPU communication in the UPC programming model.
This work distinguishes itself from other related studies in the discussion of
designs for GPU-aware communication and their performance in a message-driven runtime system
and multiple parallel programming models built on top of it, utilizing a state-of-the-art
communication library, UCX.

It should be noted that this work is an extension of~\cite{ashes21-choi},
introducing the Channel API in the Charm++ runtime system which is designed to avoid the overheads from
the GPU Messaging API. The Channel API no longer needs a host-side metadata message and
can post the receive for the incoming data as soon as the user calls a receive primitive.

\section{Conclusion}\label{sec:conclusion}

In this work, we have discussed the importance of GPU-aware communication in today's GPU-accelerated
supercomputers, and the associated technologies that are involved in supporting
direct GPU data transfers for several parallel programming models: Charm++, AMPI,
and Charm4py. We leverage the capabilities of the UCX library to implement an extension
to the UCX machine layer in the Charm++ runtime system, providing a performance-portable
communication layer for the Charm++ family of parallel programming models.
With the GPU Messaging API in Charm++, we are able to retain the semantics of message-driven execution
while demonstrating substantial performance improvements. 
We also discuss the design of the Channel API in Charm++, which deviates from message-driven
execution to provide data-only communication that can be useful for certain types of
applications. The Channel API demonstrates superior performance to the GPU Messaging API
due to its simpler design and a more direct interface to the underlying UCX library.
Our GPU-aware communication mechanisms demonstrate superior performance over the host-staging methods
in micro-benchmarks adapted from the OSU benchmark suite, as well as a proxy application
that represents a widely used stencil algorithm.

With GPU-aware communication support in place for the Charm++ ecosystem,
we plan to incorporate computation-communication overlap with overdecomposition~\cite{espm220-choi}
to minimize communication overheads on modern GPU systems.
We also plan on supporting collective communication
of GPU data, using this work as the basis to translate collective communication
primitives into GPU-aware point-to-point calls.

While UCX proves to be an effective framework for universally
accelerating GPU communication, there is still room for performance improvement
as indicated by the differences between AMPI and OpenMPI.
One of the potential areas of improvement is GPU support in the Active Messages
API of UCX, which could better fit the message-driven execution model of Charm++.
Another is replacing the use of the GPU Messaging API for AMPI and Charm4py
with the new Channel API,
which would eliminate the need to delay the posting of the receive for GPU data
until the arrival of the metadata message.


\section*{Acknowledgment}

\newcommand\blfootnote[1]{%
	\begingroup
	\renewcommand\thefootnote{}\footnote{#1}%
	\addtocounter{footnote}{-1}%
	\endgroup
}

We thank the UCX developer team, including Akshay Venkatesh, Devendar Bureddy, and Yossi Itigin
for their assistance with technical issues on the Summit supercomputer.

This work was performed under the auspices of the U.S. Department
of Energy (DOE) by Lawrence Livermore National Laboratory
under Contract DE-AC52-07NA27344 (LLNL-JRNL-826064).

This
research was supported by the Exascale Computing Project (17-SC-20-SC),
a collaborative effort of the U.S. DOE Office of Science and
the National Nuclear Security Administration.

This research used resources of the Oak Ridge Leadership Computing
Facility at the Oak Ridge National Laboratory, which is
supported by the Office of Science of the U.S. DOE under Contract
No. DE-AC05-00OR22725.






\bibliographystyle{elsarticle-num}
\bibliography{ref}

\end{document}